\documentclass[useAMS,usenatbib]{mnras}
\usepackage{amsmath}
\usepackage{amssymb}
\usepackage{float}
\usepackage{graphicx}
\usepackage{subcaption}
\usepackage{mathptmx}
\usepackage{newtxtext,newtxmath}
\usepackage{multirow}

\newcommand{\msun}{M$_{\sun}$}
\newcommand{\zsun}{Z$_{\sun}$}
\newcommand{\mseed}[1]{$M_{\mathrm{seed}}=10^{#1}\,\mathrm{M}_{\sun}$}
\newcommand{\ramses}{\textsc{ramses}}

\newcommand{\unit}[1]{\, \mathrm{#1}}
\newcommand{\sub}[1]{_{\mathrm{#1}}}

\newcommand{\kms}{km\,s$^{-1}$}

\graphicspath{{./figures_feedback/}}
 
\defcitealias{Biernacki2017}{BTB17}

\def\equationautorefname~#1\null{Eq.~(#1)\null}
\title[Combined AGN and SN feedbacks launching outflows]{The combined effect of AGN and supernovae feedback in launching massive molecular outflows in high-redshift galaxies}

\author[P. Biernacki \& R. Teyssier] {Pawel Biernacki\thanks{biernack@physik.uzh.ch} and Romain Teyssier\\ \\
  {Centre for Theoretical Astrophysics and Cosmology, Institute for Computational Science, University of Zurich,} \\
  {Winterthurerstrasse 190, 8057 Zurich, Switzerland} \\
}

\begin{document}
\maketitle
\begin{abstract}
We have recently improved our model of active galactic nucleus (AGN) by attaching the supermassive black hole (SMBH) to a massive nuclear star cluster (NSC). Here we study the effects of this new model in massive, gas-rich galaxies with several simulations of different feedback recipes with the hydrodynamics code RAMSES. These simulations are compared to a reference simulation without any feedback, in which the cooling halo gas is quickly consumed in a burst of star formation. In the presence of strong supernovae (SN) feedback, we observe the formation of a galactic fountain that regulates star formation over a longer period, but without halting it. If only AGN feedback is considered, as soon as the SMBH reaches a critical mass, strong outflows of hot gas are launched and prevent the cooling halo gas from reaching the disk, thus efficiently halting star formation, leading to the so-called "quenching". If both feedback mechanisms act in tandem, we observe a non-linear coupling, in the sense that the dense gas in the supernovae-powered galactic fountain is propelled by the hot outflow powered by the AGN at much larger radii than without AGN. We argue that these particular outflows are able to unbind dense gas from the galactic halo, thanks to the combined effect of SN and AGN feedback. We speculate that this mechanism occurs at the end of the fast growing phase of SMBH, and is at the origin of the dense molecular outflows observed in many massive high-redshift galaxies.
\end{abstract}

\section{Introduction}\label{sec:intro}

A successful model of galaxy formation must reproduce both the observed stellar masses and spatial distributions. 
Current star formation recipe and their associated feedback mechanisms appear to be able to regulate the stellar content in small mass halos ($M\sub{halo}<5\times10^{11}$\,\msun), but less so in the most massive galaxies \citep{Shankar2006, Dave2011, Moster2013, Behroozi2013}.
Supermassive black holes (SMBHs) are good candidates to quench star formation in early-type galaxies \citep{Nandra2007, Schawinski2007, Fabian2012, Yesuf2014, Cheung2016}, as the energy released by active galactic nuclei (AGN) could be large enough to unbind significant amounts of star-forming gas.

SMBHs are ubiquitous elements of galactic environments at all redshifts \citep[see e.g. the review by][]{Cattaneo2009} 
-- starting with the Milky Way \citep{Schodel2002,Gillessen2009}, to galaxy groups and clusters \citep[e.g.][]{Magorrian1998}, 
up to luminous $z>6$ quasars \citep{Fan2003}. 
The scaling relations between SMBH mass and its host properties, like the bulge mass or the central velocity dispersion \citep{Ferrarese2000, Laor2001, Haring2004, Tremaine2002, Gultekin2009, Kormendy2013}, indicate a strong connection between the SMBH and its host, sometimes referred to as ``coevolution''.

AGN feedback is especially important for groups and clusters of galaxies. 
Even if some properties of the intragroup medium can be explained by SN feedback alone, 
a powerful central source is necessary to really quench SF \citep[see e.g. recent work by][]{Liang2016}. 
Similar effects are seen in most massive clusters of galaxies where presence of AGN feedback is required to match observations
\citep[e.g.][]{Puchwein2008,Teyssier2011,Martizzi2013,LeBrun2014,Planelles2014,Rasia2015,Schaye2015,Hahn2017}.

Efficient SF requires a large reservoir of cold and dense gas. In order to suppress SF in large galaxies, we must reduce this reservoir dramatically. 
This can happen through two different channels.

First, we can expel this reservoir of dense molecular gas out of the galactic disk. 
This is what happens in low mass galaxies, where SN feedback leads to the production of a galactic fountain 
\citep[e.g.][]{Dave2011} and, in case of dwarf galaxies, to a strong outflow completely removing the gas
\citep[see e.g.][]{Dekel1986}. In massive haloes, however, the escape velocity is too high 
for SN feedback to play a significant role \citep{Dave2011, Zhang2012}, while AGN feedback is believed to take over.
However, a single, centrally located source cannot influence SF in the entire galactic disk,
like SN feedback does. Indeed, {as demonstrated by \cite{Roos2015}, the \emph{expulsive} feedback from the AGN has no effect on the instantaneous star formation rate (SFR) or star formation efficiency (SFE) in the galaxy, but could lead to a secular effect by reducing slowly its gas content}.

Second, quenching of star formation can be the result of cutting external gas supplies, so that the
existing dense gas reservoir is consumed by the local SF and not replenished.
In other words, if gas outside the disk gets expelled from the halo or stopped from being accreted, 
then it cannot contribute to star formation. 
This \emph{preventive} feedback has been identified in two viable mechanisms for AGN feedback in massive halos: 
1) the so-called quasar mode, for which giant outflows halt the global inflow from filamentary accretion, and 
2) the so-called radio mode, for which narrow radio-loud jets maintain the halo gas in hydrostatic equilibrium 
by balancing cooling \citep{Bruggen2002, Pizzolato2005, Sijacki2007, Ciotti2010, Gaspari2013, Choi2015, Li2015, Hopkins2016} 

Regarding the physics of outflows, observations reveal a very rich and complex picture. Hot and diffuse outflows have been seen in ultra-luminous infrared galaxies \citep[ULIRGS;][]{Sturm2011, Veilleux2013, Spoon2013}, while high-$z$ observations of massive QSOs \citep[e.g.][]{Cicone2014} display outflows consisting of cold, molecular gas moving with high-velocities. As shown recently by \citet{Costa2015}, at $z> 6$ molecular outflows can be explained by hot, AGN-driven gas which cools due to mixing with metal-enriched SN-powered gas and possibly an interaction with cold streams.

Recently, Chapman et al. (submitted) reported a molecular outflow with several $10^{10}$ \msun{} of gas moving with velocities reaching 1500\,\kms{} at $z=2.85$. 
These extreme outflows pose a severe challenge for galaxy formation models. 
Because they cannot be driven by stellar winds or SN explosions due to their low energetics,
AGN feedback appears as a natural explanation \citep{Angles-Alcazar2014, Choi2015}. Indeed, hot, low-density outflows associated with Broad Absorption Lines seen in connection to AGN activity, possibly driven by radiation pressure on dust grains, can reach outflow velocities up to 30'000\,km/s \citep{Scoville1995, Thompson2015}, thus providing the necessary kinetic energy to unbind the galactic gas.

Understanding these two very different mechanisms (hot versus cold outflows) and their possible interplay is still a matter of active research. 
Isolated AGN feedback does not produce gas outflow morphologies as seen in observations, 
while SN feedback can produce cold, dense outflows, but they remain bound to the disk. 
Some models have been proposed featuring a competition between these two processes. 
\cite{Dubois2015} showed that SMBHs cannot grow significantly in the presence of fervent SF and efficient SN feedback, 
which is the case at the peak of SF around $z=2-3$, and that fast SMBH growth is allowed only when there is a large enough galactic bulge.
In \citet[][hereafter \citetalias{Biernacki2017}]{Biernacki2017}, we have also shown that, 
if the SMBH is hosted by a nuclear star cluster (NSC), 
it can grow efficiently, creating the conditions for a possible cooperation between AGN feedback in the nuclear region and SN feedback in the extended disk.

In this work, we report on the effects of AGN feedback on regulating SF in the galactic environment and launching strong gas outflows within the halo. 
We study specifically the case of gas-rich, massive, high-$z$ galaxies, progenitors of the massive ellipticals we see today at the heart of groups and clusters. 
Our setup, due to our rather high numerical resolution, allows us to explore the interplay between the SMBH and the interstellar medium (ISM), as well as the effect of AGN feedback on the galactic corona. 
In \citetalias{Biernacki2017}, we have presented an improved sink particle implementation for SMBH formation and evolution. This new model is used here to study the effect of the SMBH to the host galaxy and halo. 
In \autoref{sec:setup}, we briefly summarise the details of our model and present the numerical setup. 
In \autoref{sec:sfr}, we discuss the evolution of the SF and its quenching by AGN feedback,
demonstrating that it acts as a \emph{preventive} mechanism, 
while in \autoref{sec:outflows}, we focus on the properties of the  gas outflows, focusing on their velocities. \autoref{sec:morphology} is devoted to the analysis of gas morphologies in our simulations. 
In \autoref{sec:discussion} we are discussing our results in the context of current observations of molecular outflows
and how they can be applied to large-scale cosmological simulations. 
We summarise our findings in \autoref{sec:summary}.


\section{Numerical setup}\label{sec:setup}

The simulations discussed in this paper have been already presented in detail in \citetalias{Biernacki2017}.
We only recall the aspects of our numerical setup that are particularly relevant to this study.

We have run our simulations with the adaptive mesh refinement (AMR) code \ramses{} \citep{Teyssier2002}. 
Gas hydrodynamics is based on solving the Euler equations with a second-order, unspilt Godunov scheme. 
Stars and dark matter are modelled using collisionless particles, that are evolved with an adaptive Particle-Mesh $N$-body solver.

Our initial conditions are designed to mimic a typical high redshift galaxy. 
We start with an isolated, gas-rich, slowly rotating dark matter halo of mass $2\times10^{12}$ \msun{}, with a spin parameter of $0.04$.
We sampled it with one million dark matter particles. The halo profile follows a truncated NFW profile with a concentration parameter $c=10$. 
Gas in the halo is initially in hydrostatic equilibrium and follows the same NFW profile. 
The parameters of our halo are the followings: the circular velocity is $V\sub{200}=160\,\mathrm{km\,s}^{-1}$, 
the viral radius of $R\sub{200}=230\,\mathrm{kpc}$, and the halo truncation radius at $514\,\mathrm{kpc}$. 
This particular \ramses{} setup was introduced first in \citet{Teyssier2013}.

In our simulations, gas cooling for gas hotter than $10^4\,\mathrm{K}$ follows the cooling function of \citet{Sutherland1993}, 
which accounts for radiative cooling of H, He, as well as a standard mixture of metal. 
For lower temperatures, we consider only fine-structure metal cooling following the cooling function of \citet{Rosen1995}. 
The evolution of metallicity is modelled using a passive scalar, which is advected with the flow. 
We adopted an initial metallicity $Z\sub{ini}=0.05$ Z$\sub{\sun}$, where the solar metallicity was set to a metal mass fraction of Z$\sub{\sun}=0.02$.

In order to minimise numerical problems due to our limited spatial resolution of $\Delta x\sub{min} = 78\,\mathrm{pc}$,
we use a temperature floor 
\begin{equation}
T\sub{floor}=T\sub{*}\left(\frac{n\sub{\mathrm{H}}}{n\sub{*}}\right)^{\Gamma-1}\label{eq:press_floor}
\end{equation} with a critical gas number density $n\sub{*}=9\,\mathrm{cm}^{-3}$, a critical temperature $T\sub{*}=2\times10^3\,\mathrm{K}$, and $\Gamma=2$.
Our star formation prescription follows the method of \citep{Rasera2006}, which stochastically spawns stellar particles from a Poisson distribution 
using a Schmidt law if the gas density in the cell exceeds $n\sub{*}=9\,\mathrm{H/cm}^{3}$. 
The efficiency with which stars are formed is set to be $\epsilon_*=0.01$, based on values measured in local molecular clouds \citep{Krumholz2007}. 
We model supernovae explosions assuming that only 10\% of the stellar mass goes supernovae and that a single supernova injects $10^{51}$~ergs. 
Furthermore, we assume that each 10~\msun{} of ejecta contains 1~\msun{} of metals. 
We boost the efficiency of our SN explosions by grouping stars stochastically in clusters of mass $10^8$~\msun{}. 
In order to overcome the overcooling problem of supernovae feedback due to our limited resolution, 
we use a non-thermal energy variable that dissipates over a 10 Myr timescale \citep[see][for details]{Teyssier2013}.

Supermasive black holes (SMBHs) are modelled with our new sink particle algorithm \citep{Bate1995, Krumholz2004, Bleuler2014, Biernacki2017}. 
We allow for only one sink to form in our simulations. The sink formation site is identified on the fly with the clump finder \textsc{phew} \citep{Bleuler2015} 
as the first massive enough gas clump. This is usually, but not always, at the centre of the galaxy. 
The adopted initial SMBH mass (the seed mass) is a free parameter in our simulations and spans the range $10^5$~\msun{} to $10^8$\,\msun{}. 
SMBH accretes according to the Bondi-Hoyle-Lyttleton \citep[][or Bondi for short]{Hoyle1939, Bondi1944, Bondi1952} accretion
\begin{equation}\label{eq:bondi}
\dot{M}\sub{Bondi} = \frac{4\pi \rho\sub{\infty} G^2 M^2\sub{SMBH}}{\left(c\sub{s}^2/\beta\sub{boost}+v\sub{rel}^2\right)^{3/2}}
\end{equation}
where $G$ is the gravitational constant, $\rho\sub{\infty}$ is the density at infinity in the classical Bondi accretion solution \citep[for more details see][]{Krumholz2004}, 
$M\sub{sink}$ is the mass of the SMBH. We boost the accretion rate by reducing the average sound speed $c\sub{s}$ in the sink's vicinity 
with $\beta\sub{boost}$ \citep[as in][]{Booth2009}. If we set the relative velocity $v\sub{rel}=0$, then we recover exactly the solution originally proposed by \cite{Booth2009}.
	
The dynamical evolution of the sink particle is modelled using a direct summation method for the gravity between the sink and the matter. 
It is more accurate than the Particle Mesh method in case of very massive SMBHs dominating the local gravitational potential. 
Furthermore, we include an additional drag force due to accretion, which leads to additional momentum exchange between the sink and the surrounding gas. 
This is performed by requiring that 1) the centre of mass of the sink-gas system remains fixed during the accretion and 2) the total linear momentum is conserved.

In a subset of our simulations, we also include a simple model of coevolution of the SMBH and a host nuclear star cluster (NSC). 
Here, the sink particle mass is the sum of the two components $M\sub{sink}=M\sub{SMBH}+M\sub{NSC}$ and is the mass used in all the gravity calculations. 
SMBH grows with the  Eddington-limited Bondi rate (\autoref{eq:bondi}), while for the growth of NSC, we used a simple model for which $\dot{M}\sub{acc, NSC}=100\dot{M}\sub{acc, SMBH}$. 
As shown in \citet{Biernacki2017}, this prescription allows us to solve the problem of wandering SMBHs by locking them within a central massive stellar component 
- either a NSC \citep[as hinted by observations of e.g.][]{Seth2008} or a bulge \citep[e.g.][]{Silk1998,Magorrian1998}.

Finally, feedback from the AGN is modelled with a simple thermal energy injection in the vicinity of the sink,
within the radius of $R\sub{sink}=4\Delta x\sub{min}$ from the sink position, where $ \Delta x\sub{min}$ is the size of the smallest resolution element. 
The luminosity of the AGN is calculated as
\begin{equation}
L\sub{AGN} = \epsilon\sub{c}\dot{M}\sub{acc}\epsilon\sub{r}c^2,
\end{equation}
with $\epsilon\sub{r}=0.1$ being the accretion disk radiative efficiency \citep[in the so-called quasar mode;][]{Shakura1973} 
and $\epsilon\sub{c}=0.15$ representing the hydrodynamic coupling efficiency, which was calibrated in previous works done with the \ramses{} code 
\citep{Teyssier2011, Dubois2012,Gabor2013}. In \autoref{tab:summary} we summarise all simulation parameters used in this work, 
including a reference to the section in which they are first discussed.
\begin{table}
\begin{center}
\caption{Summary of simulations discussed in this work. Columns: (1) subsection in which the simulations are first analysed; (2) SN feedback modelling (yes/no); (3) SMBH seed mass, if present; (4) final stellar mass of the galaxy (i.e. after 1.5~Gyr).}
\begin{tabular}{cccc}
\hline
\hline
Section  & SN feedback & $m\sub{seed}$ [\msun] & M$_*$ [$10^{10}$\,\msun]\\
(1) & (2) & (3) & (4)\\
\hline\hline
\multirow{5}{*}{\ref{ssec:nosn}} 	& \multirow{5}{*}{no} 	& -- 		& 8.5\\
		  				 	&  					& $10^5$ & 8.5\\
		  				 	&  					& $10^6$ & 5.5\\
							&  					& $10^7$ & 4.7\\
							&  					& $10^8$ & 4.1\\
\hline
\multirow{5}{*}{\ref{ssec:sn+agn}}	& \multirow{5}{*}{yes}  	& -- 		& 7.4\\
							&  					& $10^5$ & 7.5\\
							&  					& $10^6$ & 4.9\\
							&  					& $10^7$ & 4.6\\
							&  					& $10^8$ & 3.9\\
\hline\hline
\end{tabular}\label{tab:summary}
\end{center}
\end{table}


\section{Mass accretion and star formation}\label{sec:sfr}

\begin{figure*}
\centering
\includegraphics[width=\textwidth]{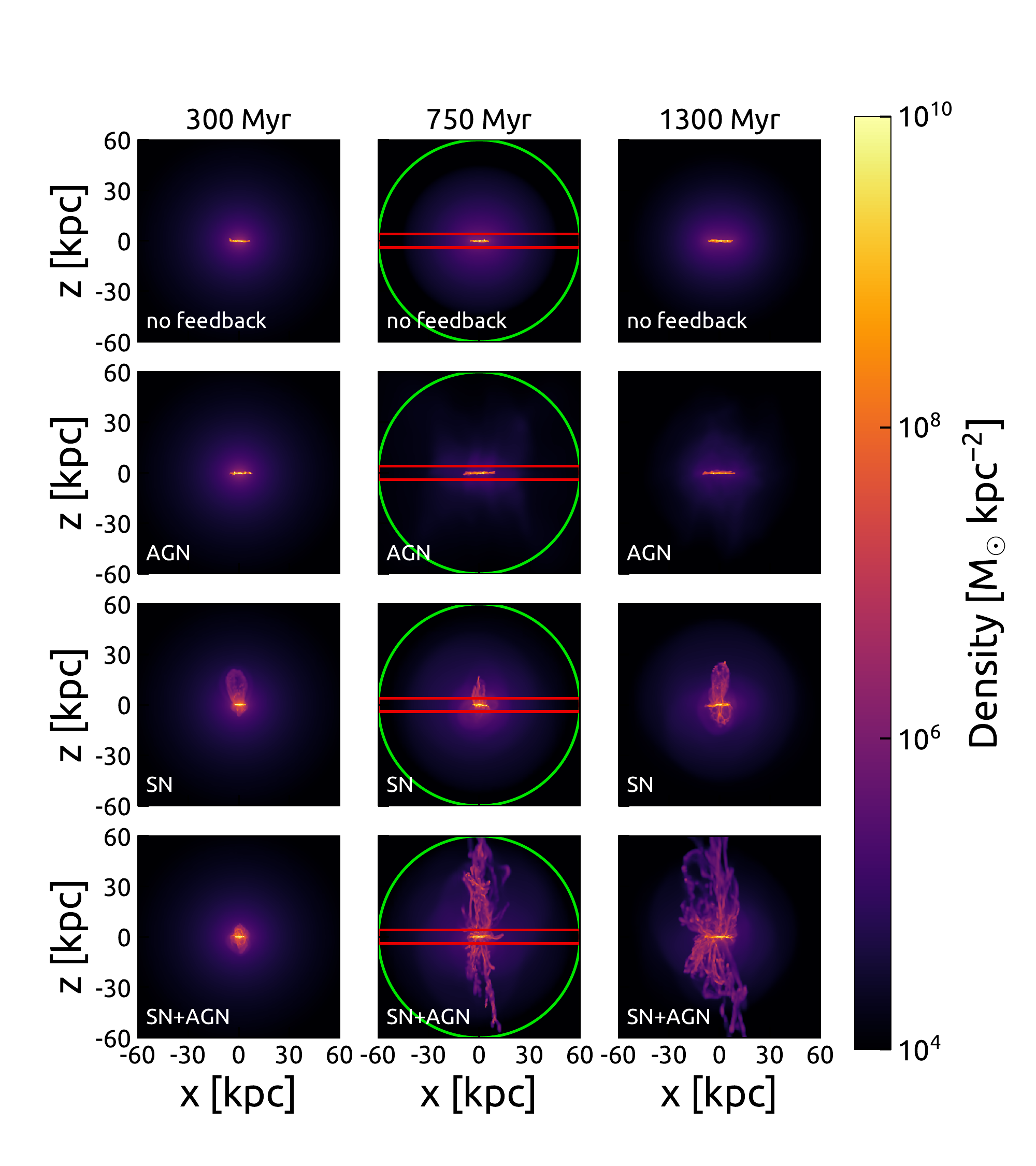}
\caption{Gas surface density for four feedback models (top to bottom: no feedback, AGN, SN, SN+AGN) at three different times (300, 750 and 1300 Myr; left to right). In all cases galaxy is shown edge-on. \mseed{6} in simulations with AGN feedback modelled. Green circles mark gas selected on \autoref{fig:histograms}, while the gas from red rectangle has be excluded from the analysis.}\label{fig:rhorho}
\end{figure*}

In this Section, we discuss the results of our simulations in term of SFR and how it can be impacted by AGN feedback. 
In \autoref{fig:rhorho}, we show the gas surface density of the disk in an edge-on projection for four of our simulations: 
no feedback, AGN feedback-only, SN feedback-only, and SN+AGN feedback (from top to bottom) at three different times 300, 750 and 1300 Myr (from left to right). 
We choose these particular times during different important epochs -  the earliest time corresponds to the epoch when SF is at its highest, 
while the SMBH is still growing and did not impact the host galaxy, the intermediate time corresponds to the epoch of AGN outflow launching, 
while the latest snapshot shows the final state of the galaxy.

\subsection{Simulations without supernova feedback}\label{ssec:nosn}

In order to quantify the impact of AGN feedback on the star formation history, we always compare to a reference model in which both AGN and SN feedbacks are not modelled 
-- the `no feedback' simulation (\autoref{fig:rhorho}, first row). 
It can be seen that the extended initial gaseous halo has settled into a centrifugally supported disk. 
This large reservoir of gas is slowly consumed by star formation. 
In \autoref{fig:sfr_agn} we plot this reference star formation history with a black dotted line. 
The peak of star formation happens roughly 300 Myr after the beginning of the simulation and then decays exponentially.
This is because fresh gas infall from the outer halo is also slowly decaying; the gas has to be brought to the disk from increasingly larger radii. 
In \autoref{fig:inflow_agn}, we show the mass accretion rate, measured using 1-kpc-thick shells placed at 20 and 50\,kpc from the centre of the halo 
(top and bottom left, respectively; black dotted line). Clearly, the SFR correlates well with the inflow of the gas from the extended halo. 
At later times, the SFR reaches its lowest value around 30 to 40~\msun{}/yr, which is precisely the residual mass accretion rate from the halo we measure at 20\,kpc. 
The characteristic mass accretion profile is related to the shape of the NFW profile we have adopted for our initial conditions. 
The sharp fall off after 800 Myr (at 50\,kpc) is related to the truncation radius of our halo.

The second row of \autoref{fig:rhorho} presents the side-on gas surface density for one of our simulations with AGN feedback. 
One can see that the gas distrbution is very similar to the `no feedback' simulation, with however significantly less gas in the halo.
In \autoref{fig:sfr_agn}, we plot the SFR for our simulations with AGN feedback, in which we varied the initial seed mass. 
In \citetalias{Biernacki2017}, we showed, that the time it takes for the SMBH to reach its self-regulated, final mass is directly related to its seed mass. 
We see then that quenching of star formation occurs precisely when the SMBH reaches its maximum mass. 
Initially, the SFR is only slightly reduced due to gas being consumed by the SMBH in its vicinity. 
Once the SMBH reaches it self-regulated mass -- for which it is able to deposit enough energy to overcome cooling -- AGN feedback drives powerful outflows. 
Those halt the infall of fresh gas from the halo, effectively leading to the starvation of the disk - the halo gas is prevented from replenishing it with star-forming gas. 
This naturally occurs earlier for larger SMBH initial seed masses. 
As an effect of this quenching, SFR settles in a very  inefficient state, around $\sim\,10$\,\msun{}/yr, 
which is a factor of four less compared to simulation without AGN feedback (black dotted line on \autoref{fig:sfr_agn}). 
Traditionally, quenching refers to a state where almost no new stars are formed and the galaxy slowly turns `red'. This is not what we obtain here, 
but the total stellar mass is significantly reduced compared to run without AGN (or equivalently with a small seed mass) -- see \autoref{tab:summary}. 
In the simulation with the smallest seed mass, $10^5$\,\msun{}, the SMBH is unable to grow and do any significant damage to the gas inflow. 

A comparison of the various mass accretion rates (\autoref{fig:inflow_agn}), measured at 20 and 50 kpc, 
reveals that AGN feedback does indeed reduce significantly the gas inflow rate, especially immediately after the first outflow is launch. 
At later times, the inflow rates (especially when measured at 50 kpc) increase again due to transverse flows parallel to the disk plane, 
bringing in the gas from outer regions. Even at late time, though, re-accreted gas cannot reach the inner disc in large quantities, 
demonstrating that AGN is able to maintain this quiescent state for a long time. This phenomenon will be discussed in more detail in \autoref{sec:outflows}.

\subsection{Simulations with supernova feedback -- SN+AGN cooperation}\label{ssec:sn+agn}

\begin{figure*}
\begin{subfigure}{0.45\textwidth}
\centering
\includegraphics[width=\columnwidth]{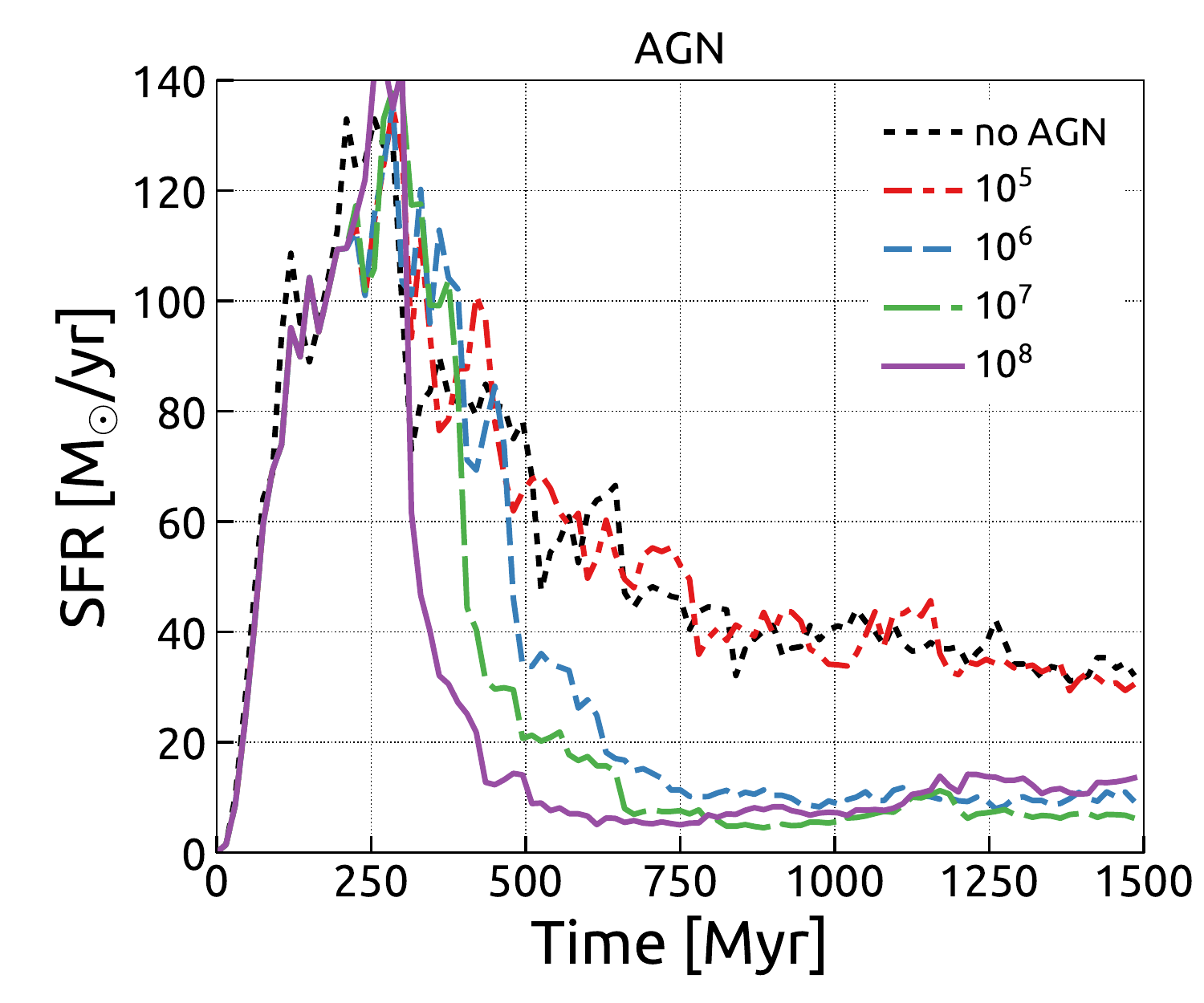}
\caption{AGN}\label{fig:sfr_agn}
\end{subfigure}
\qquad
\begin{subfigure}{0.45\textwidth}
\centering
\includegraphics[width=\columnwidth]{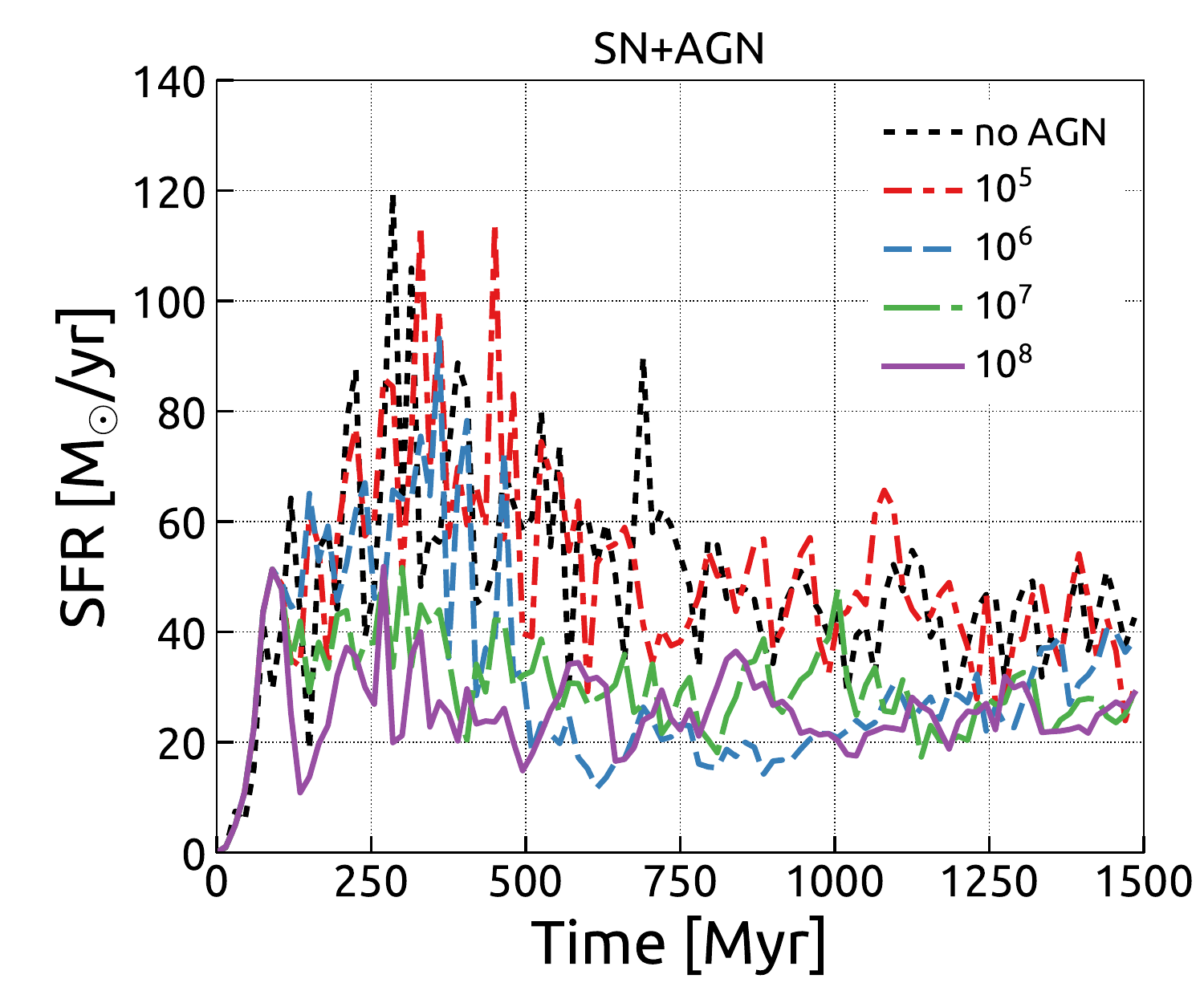}
\subcaption{SN+AGN}\label{fig:sfr_sn_nsc}
\end{subfigure}
\caption{Star formation rate (averaged over period of 15 Myr for clarity) in two sets of simulations - with AGN feedback only (left) and with both SN and AGN feedbacks (right) - for four different seed masses: $10^5$~\msun{} - red (dash-dotted), $10^6$~\msun{} - blue (short dashes), $10^7$~\msun{} - green (long dashes), $10^8$~\msun{} - purple (solid). Black dotted lines mark runs without AGN feedback.}
\label{fig:hsc_agn}
\end{figure*}

A very different picture emerges from simulations with SN feedback. We observe the onset of a galactic fountain. 
Gas inflow from the halo triggers star formation, which is then regulated by powerful gas outflows launched by SN explosions. 
These do not only remove gas, but also locally reduce the inflow from the galactic halo. This in turn leads to a reduction of the star formation and a new cycle begins. 
The resulting SFR is shown in \autoref{fig:sfr_sn_nsc} as a black dotted line. We see more scatter due to these repeated star bursts. 
Furthermore, looking at the long term evolution, we see that the peak of the SFH is lower than simulations without SN feedback and more extended.

The global star formation rate has only been slightly reduced.
In the simulations without any feedback stellar mass at the end of the runtime is $M_*=8.5\times10^{10}$\,\msun{}, 
while in the run with SN feedback this is reduced by about 10-15\% to $M_*=7.4\times10^{10}$\,\msun{} (see \autoref{tab:summary}). 
This extended, gas-rich galactic fountain  (see third row on \autoref{fig:rhorho}), 
leads to a mere redistribution of gas (and its associated star formation) to larger galactic radii \citep[see the recent work of][]{Sokolowska2016}.

In simulations with efficient SN feedback, gas accretion onto the SMBH (and thus AGN feedback) is regulated by SN feedback; 
here the final self-regulated mass of the SMBH depends mostly on the halo escape velocity \citepalias[see][for the discussion]{Biernacki2017}. 
Interestingly, the final SFR is largely independent of the SMBH seed mass. The dependence of quenching on SMBH self-regulation we see in the AGN-only runs largely disappears here.
The resulting SFR settles at $\sim20$~\msun{}/yr, thus being reduced by only a factor of two compared to the SN-only run (but still a factor of two higher than our AGN-only simulations).

\begin{figure*}
\begin{subfigure}{0.45\textwidth}
\centering
\includegraphics[width=\columnwidth]{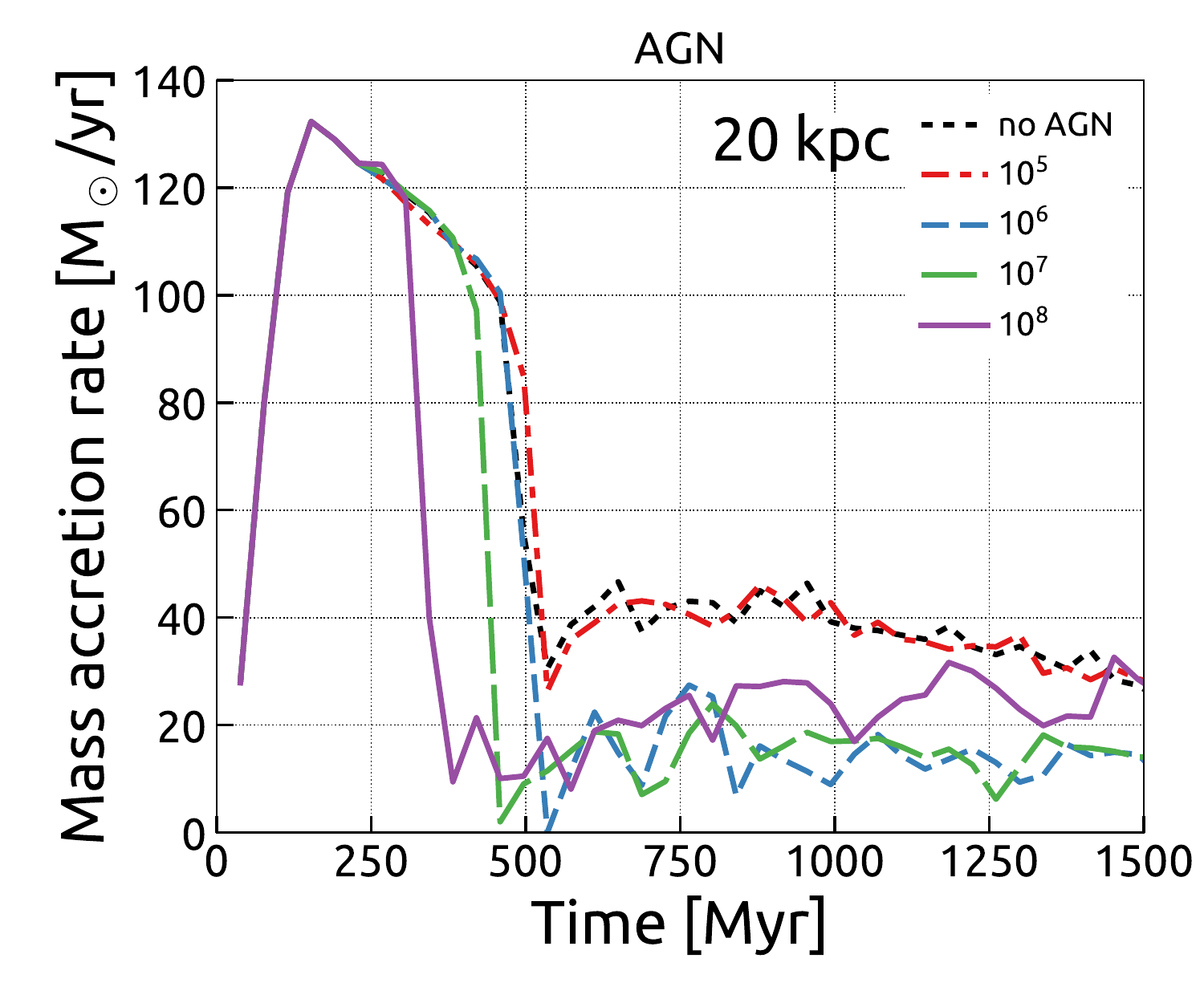}
\end{subfigure}
\qquad
\begin{subfigure}{0.45\textwidth}
\centering
\includegraphics[width=\columnwidth]{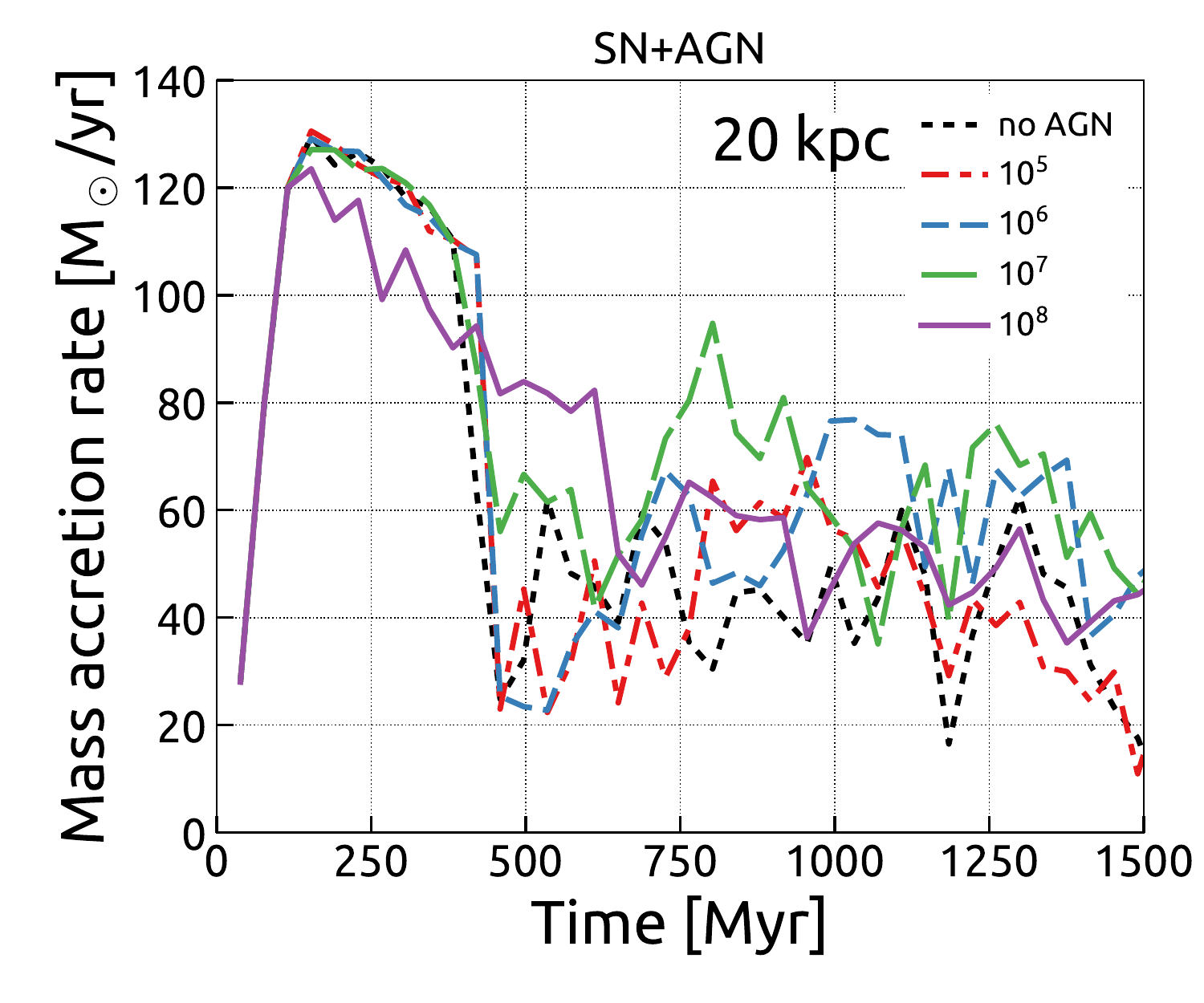}
\end{subfigure}
\\
\begin{subfigure}{0.45\textwidth}
\centering
\includegraphics[width=\columnwidth]{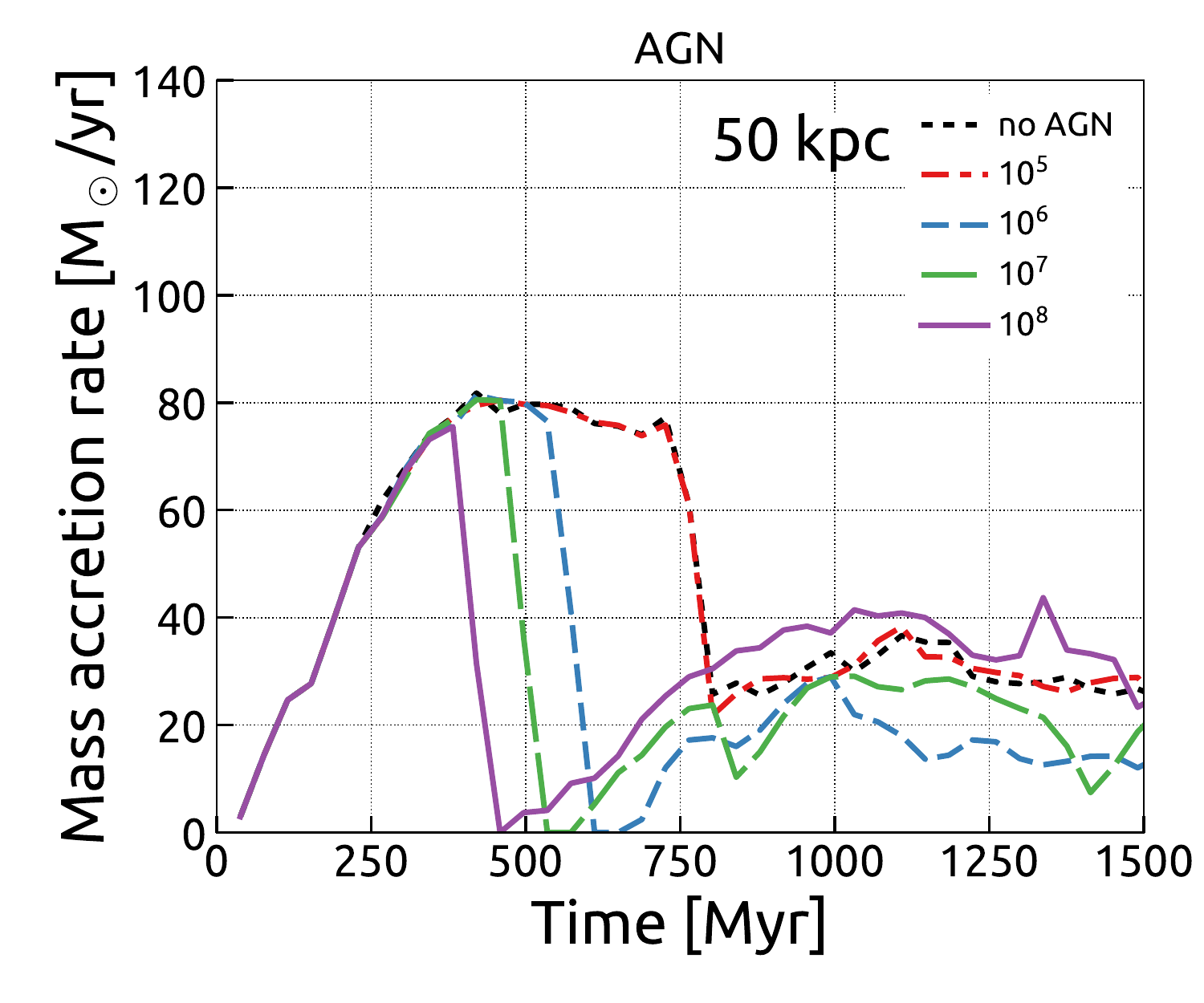}
\caption{AGN}\label{fig:inflow_agn}
\end{subfigure}
\qquad
\begin{subfigure}{0.45\textwidth}
\centering
\includegraphics[width=\columnwidth]{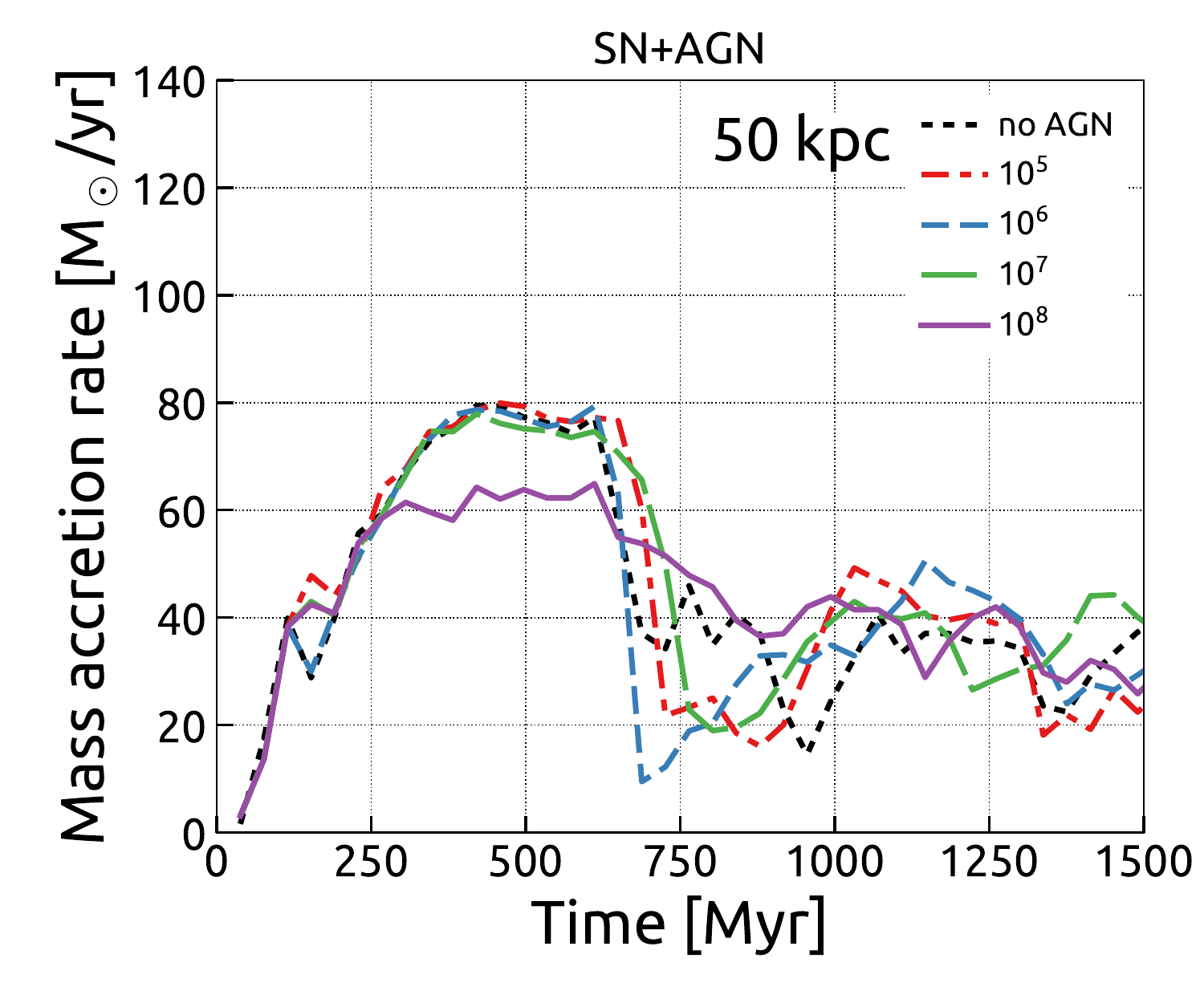}
\subcaption{SN+AGN}\label{fig:inflow_sn_nsc}
\end{subfigure}
\caption{Mass accretion rate in two sets of simulations - with AGN feedback only (left) and with both SN and AGN feedbacks (right) - for four different seed masses: $10^5$~\msun{} - red (dash-dotted), $10^6$~\msun{} - blue (short dashes), $10^7$~\msun{} - green (long dashes), and $10^8$~\msun{} - purple (solid); contrasted with runs without AGN feedback (black, dotted)). Left column: AGN only, right column: SN+AGN; top row: outflow measured through shell placed at 20 kpc from halo centre, bottom row: at 50 kpc.}
\label{fig:inflow_20+50}
\end{figure*}

The main effect of SN feedback is to inject metals which enhance the cooling of gas. As a consequence, the gas in the SN-driven galactic fountain mix with the halo gas 
and increase by a factor of two the mass inflow rate measured close to the disk (\autoref{fig:inflow_sn_nsc}, top panel), 
reducing the effect of AGN feedback on reducing the accretion of gas from the outer halo 
(perhaps with an exception of the most massive seed - bottom panel on \autoref{fig:inflow_sn_nsc}).

\section{Outflow properties}\label{sec:outflows}

\begin{figure}
\centering
\includegraphics[width=\columnwidth]{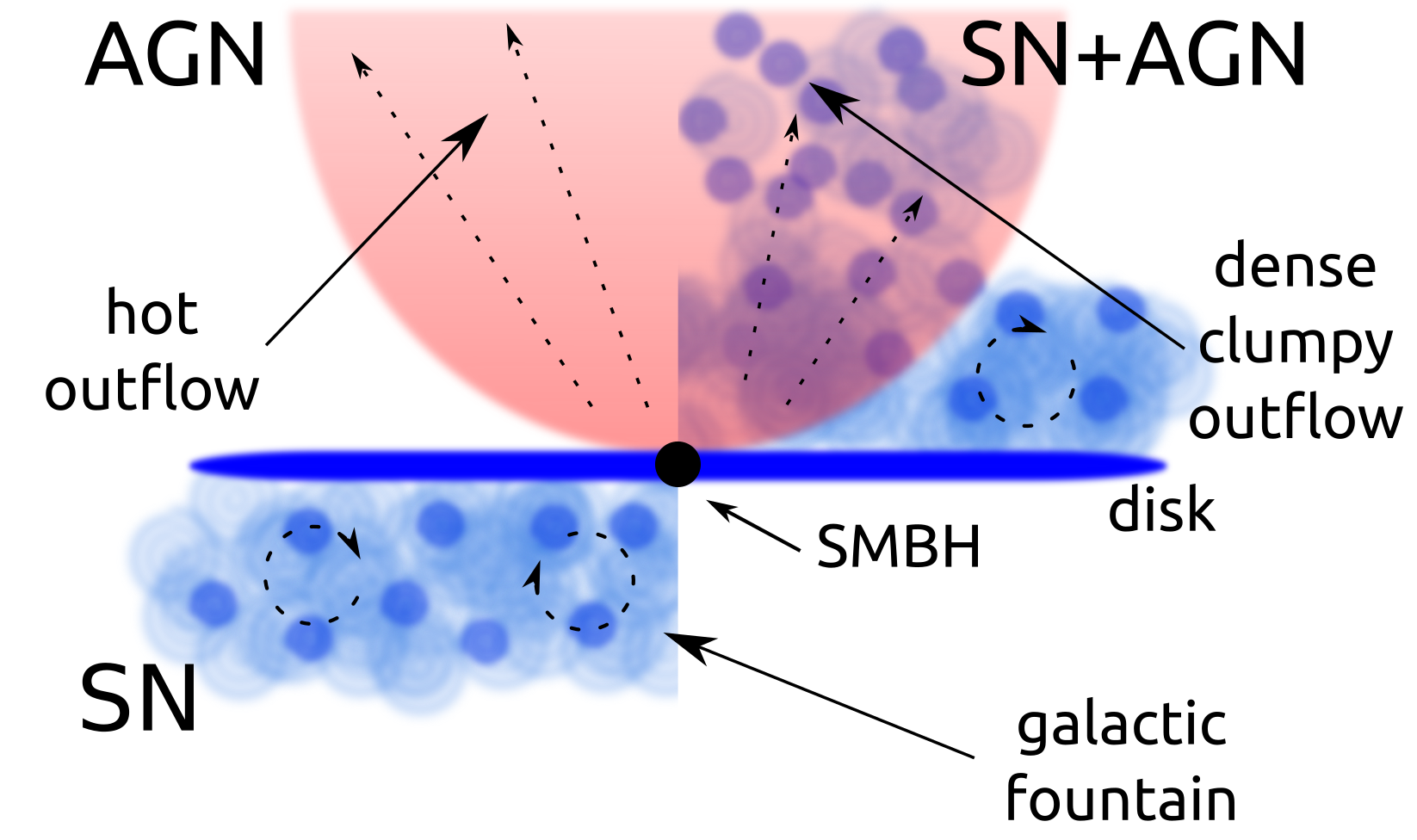}
\caption{Sketch showing different feedback modes and their impact on gas circulation. Top left: in AGN-only SMBH launches hot, diffuse outflow (pink shade); bottom left: in SN-only dense, clumpy gas forms galactic fountain in which gas is being recycled; top right: combination of to previous effects (SN+AGN) leads to dense, fast clumpy outflows which are entrained in hot outflow and escape the disk.}\label{fig:sketch}
\end{figure}

In this section, we carefully examine the properties of our AGN-driven outflows, comparing simulations without and with SN feedback. 
On \autoref{fig:sketch} we present a cartoon sketch explaining the various modes of  feedback and how they affect the properties of the outflowing gas (see also \autoref{fig:rhorho}). 
In simulations without SN feedback, the central AGN powers a strong outflow with hot and diffuse gas, 
while in simulations with SN feedback, the AGN-driven outflow interacts with a clumpy galactic fountain, 
so that cold and dense clumps are now entrained in the outflow and ejected outside the galactic corona. 

\begin{figure*}
\centering
\includegraphics[width=\textwidth]{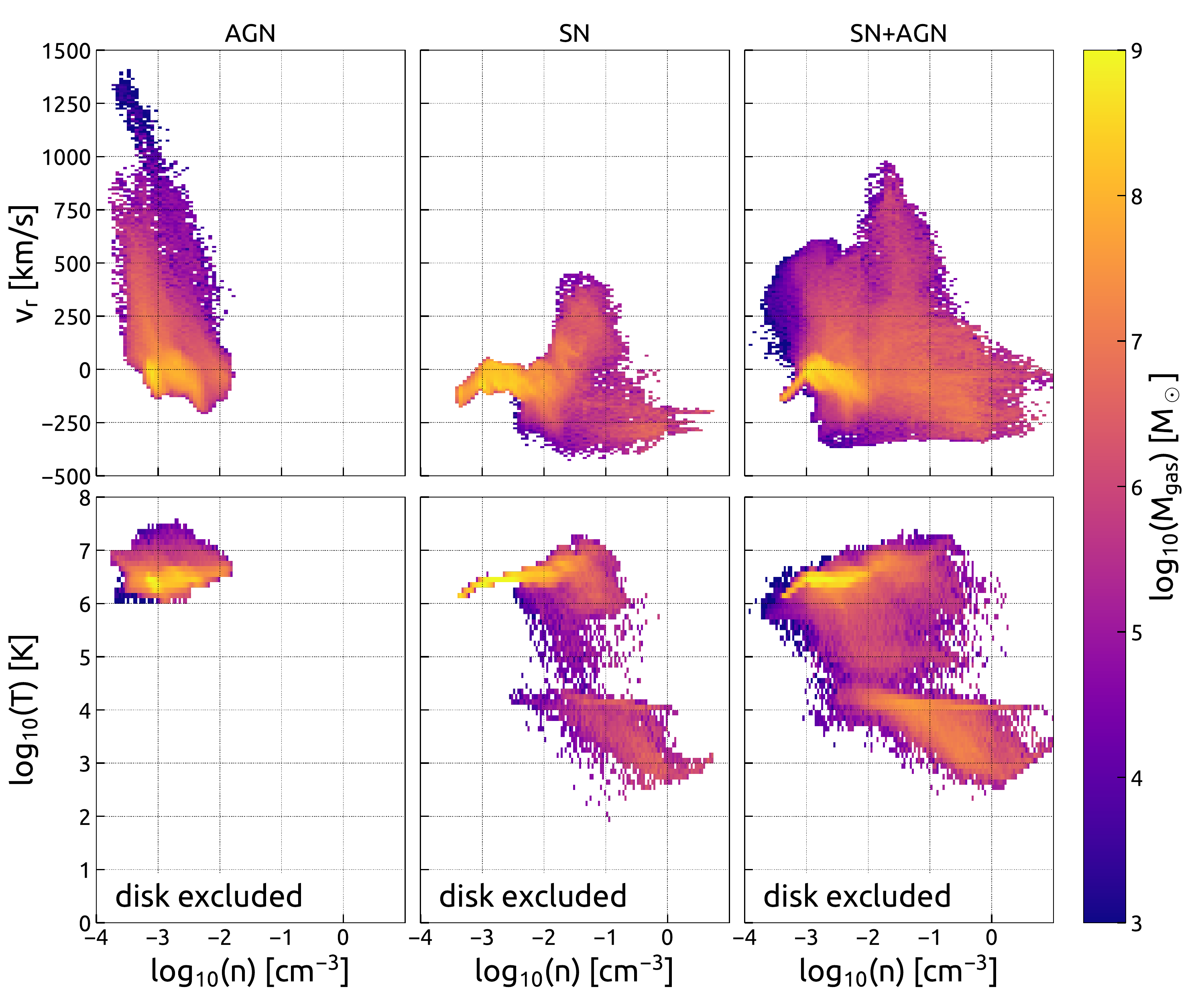}
\caption{Histograms of gas radial velocity with respect to gas density (top row) and gas density-temperature diagrams (bottom row) for three simulations with different feedback modes - AGN-only (left column), SN-only (centre), and SN+AGN (right column). We have selected only gas located at less than 60 kpc from the centre of the halo (green region on \autoref{fig:rhorho}) and excluded central disk of thickness 8 kpc (red rectangle on \autoref{fig:rhorho}). \mseed{6} in simulations with AGN feedback modelled.}
\label{fig:histograms}
\end{figure*}

\subsection{Outflow phase space diagram}\label{ssec:phase_diag}

In order to characterise the physical properties of the gas in the outflow, we restrict ourselves to a sphere of radius 60\,kpc 
(corresponding to the green region in \autoref{fig:rhorho}), 
excluding a disk  of $\pm 4$\,kpc from the disk plane
(corresponding to the red region in \autoref{fig:rhorho}).
We compute the mass fraction as a function of density and radial velocity, as well as the mass fraction as a function of density and temperature (which we discuss is more details in \autoref{ssec:temp_rho}), at 750\,Myr 
(see \autoref{fig:histograms}).
A positive radial velocity $v\sub{r}$ corresponds to \emph{outflowing} gas, while a negative value stands for \emph{inflowing} gas. 

The simulation with only AGN feedback (left column of \autoref{fig:histograms})  shows gas velocities up to 1400\,\kms{}. 
This velocity cannot be explained by buoyantly rising, AGN-driven bubbles, as the typical gas velocity in such a case would be of the order of the escape velocity 
(which is here around $700$\,\kms{}). 
As demonstrated by \citet{Costa2015}, the velocity we measure is consistent with an energy-driven wind with negligible cooling losses. 
This is indeed the case here, due to the lack of metal enrichment since no SN feedback was included in this pure AGN feedback scenario. 
For similar halo and black hole properties, the analytical model used in \citet{Costa2015} predicts maximum gas velocities in the range 1200--1800\,\kms{} (see their Figure 6). 

The simulation with only SN feedback (middle column of \autoref{fig:histograms}) features a galactic fountain with velocities up to 400\,\kms{}, 
which is less than the halo escape velocity. 
The simulation with both feedback mechanisms (right column of \autoref{fig:histograms}) shows outflowing gas with velocities typical of both feedback modes. 
Most importantly, the high velocity gas is on average one to two orders of magnitude denser than in the AGN only case. 
Here, the hot, energy-conserving outflow entrains the cold, dense gas of the fountain and accelerates it to much higher velocities. 

\subsection{Outflow mass loading factor}

We now study the temporal evolution of the mass outflow rates. 
We define the mass loading factor as the ratio of the gas \emph{outflow rate} (through a one-kpc-thick shell placed at a given radius from the centre of the halo) to the SFR of the galaxy. 
We choose 20 and 50 kpc as two representative radii - the former is at the upper edge of the galactic fountain, while the latter corresponds to a significant fraction of $R\sub{200}$ 
and captures the large-scale outflow, relevant for the entire halo.

\begin{figure*}
\begin{subfigure}{0.45\textwidth}
\centering
\includegraphics[width=\columnwidth]{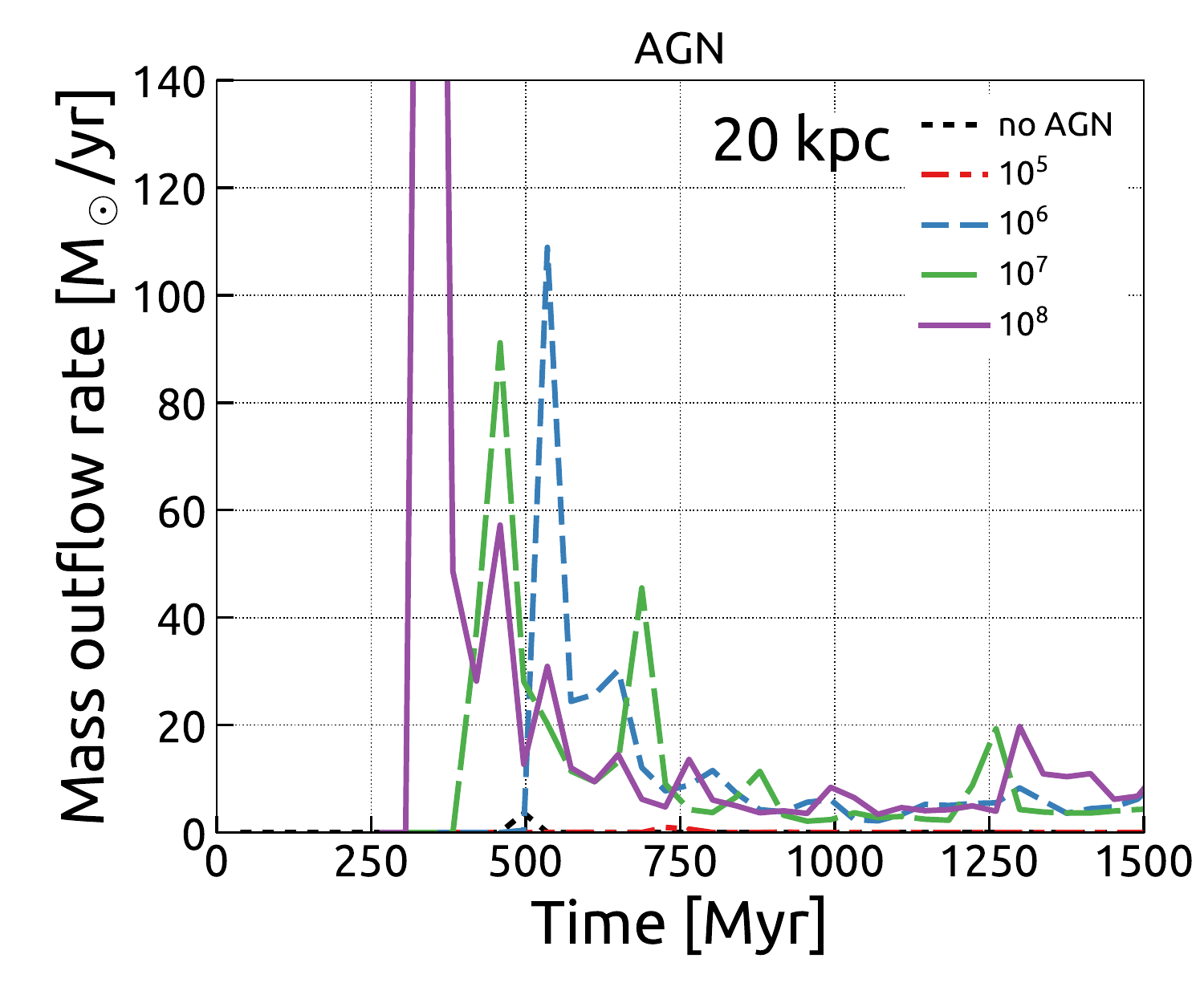}
\end{subfigure}
\qquad
\begin{subfigure}{0.45\textwidth}
\centering
\includegraphics[width=\columnwidth]{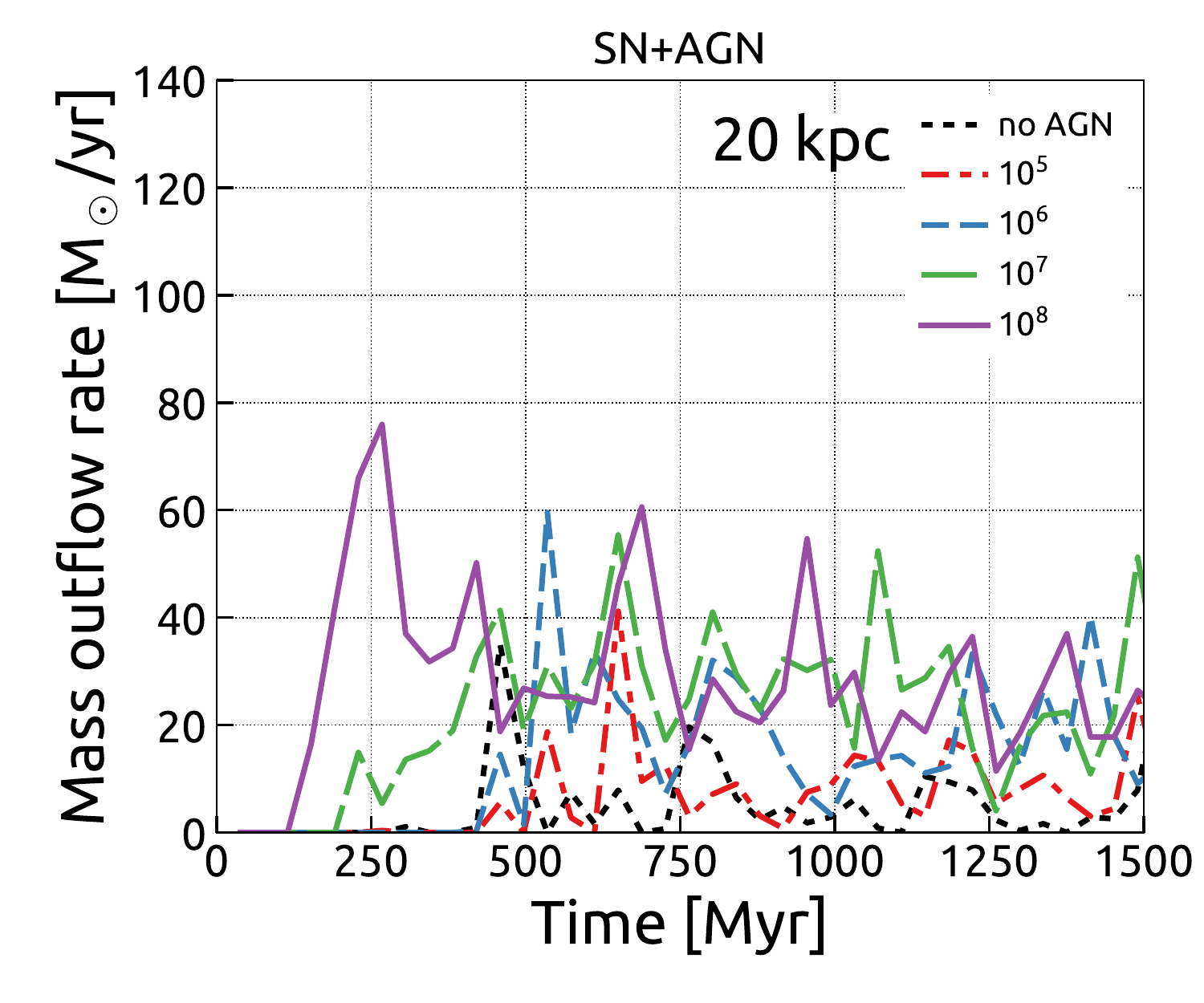}
\end{subfigure}
\\
\begin{subfigure}{0.45\textwidth}
\centering
\includegraphics[width=\columnwidth]{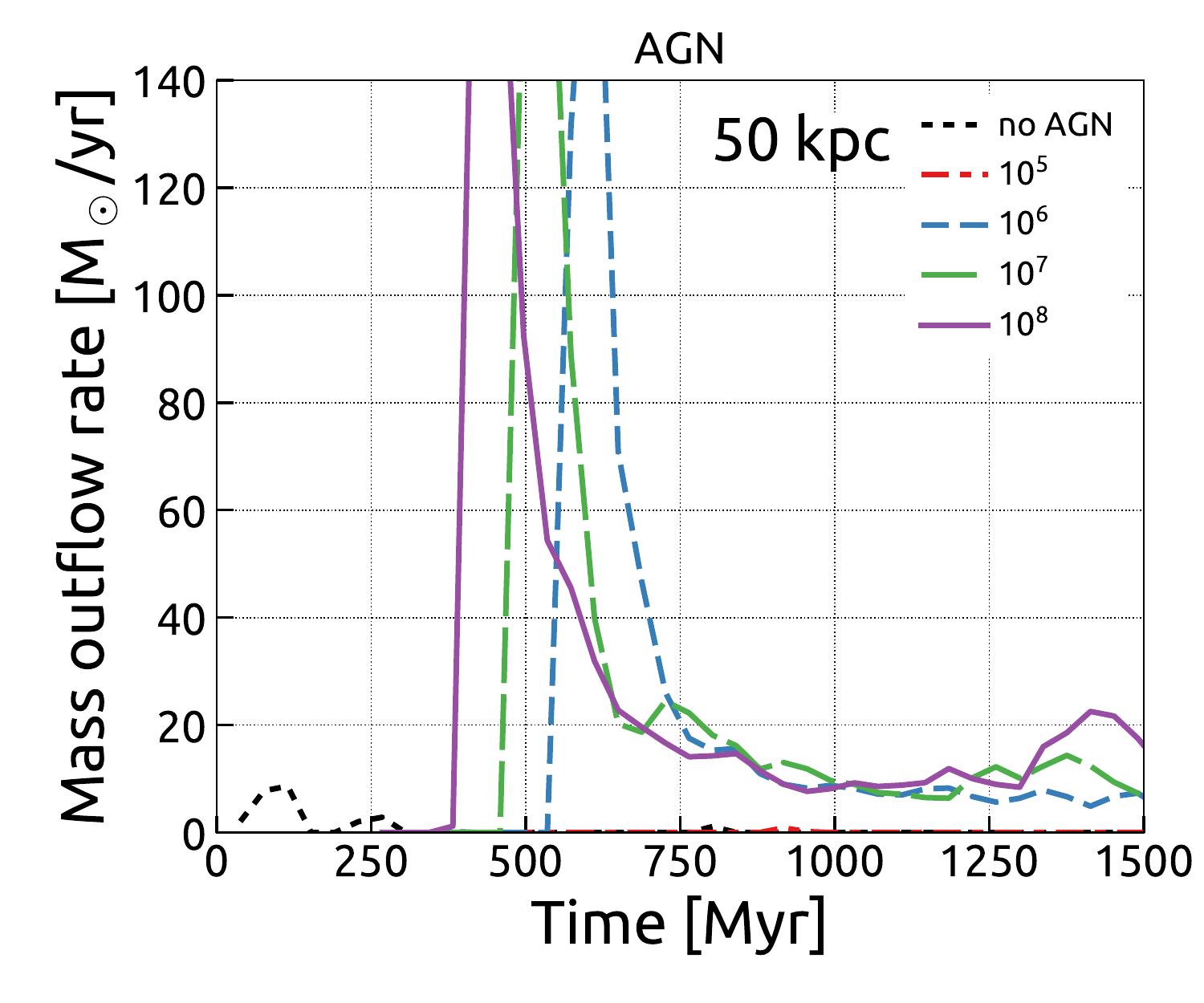}
\caption{AGN}\label{fig:outflow_agn}
\end{subfigure}
\qquad
\begin{subfigure}{0.45\textwidth}
\centering
\includegraphics[width=\columnwidth]{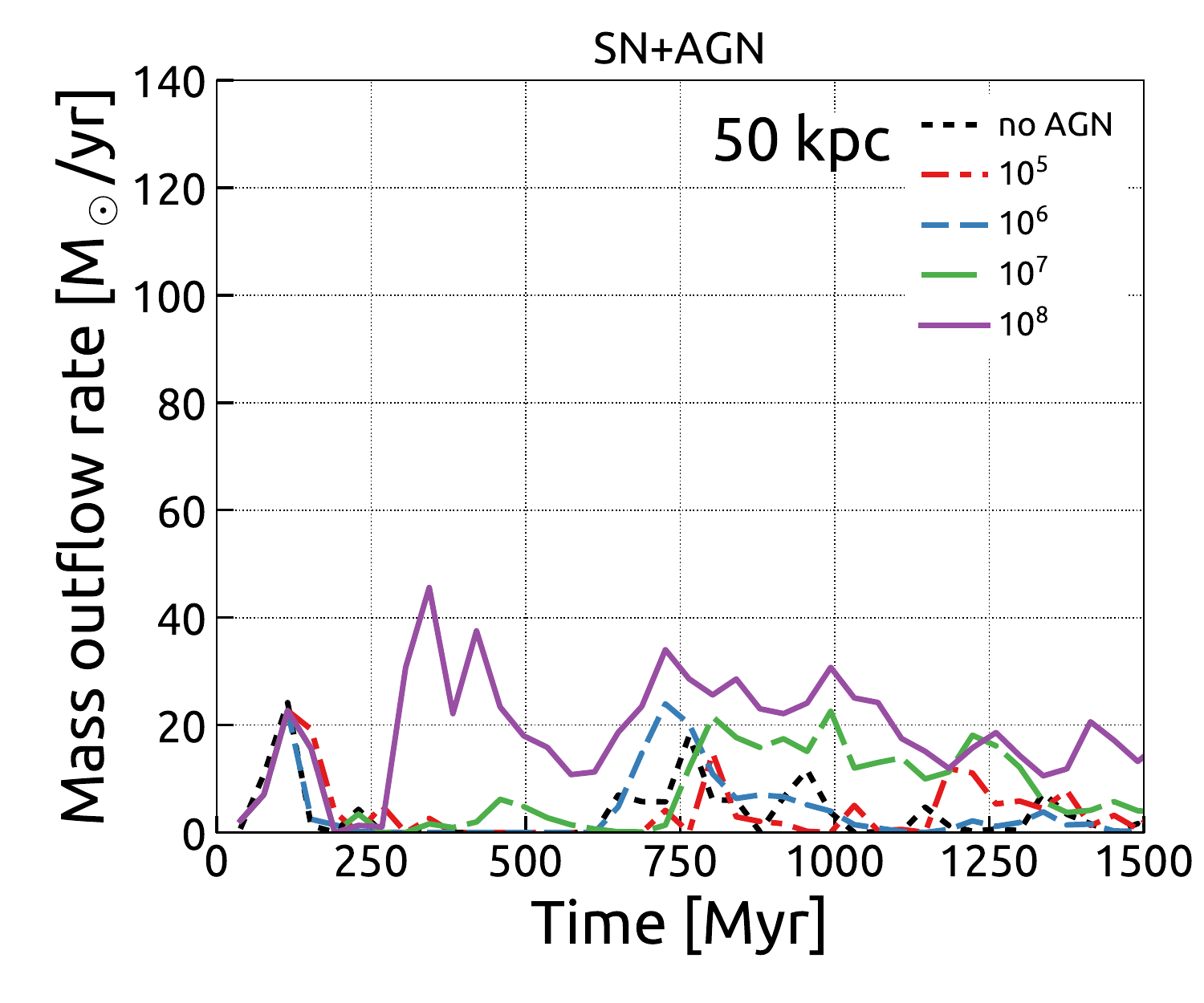}
\subcaption{SN+AGN}\label{fig:outflow_sn_nsc}
\end{subfigure}
\caption{Outflow mass in two sets of simulations - with AGN feedback only (left) and with both SN and AGN feedbacks (right) - for four different seed masses: $10^5$\,\msun{} - red (dash-dotted), $10^6$\,\msun{} - blue (short dashes), $10^7$\,\msun{} - green (long dashes), and $10^8$\,\msun{} - purple (solid); contrasted with runs without AGN feedback (black, dotted)). Left column: AGN only, right column: SN+AGN with NSC; top row: outflow measured through shell placed at 20\,kpc from halo centre, bottom row: at 50\,kpc.}
\label{fig:outflow_20+50}
\end{figure*}

In \autoref{fig:outflow_agn}, we plot the mass outflow rate of the simulations without SN feedback. 
In the reference run (black dotted line) we see no outflowing gas at all, as no feedback mechanism is present. 
The simulation with \mseed{5} is virtually identical to the `no AGN' case, since the SMBH did not grow significantly in this case. 
In the three runs with SMBH seed masses between $10^6$ and $10^8$ \msun{} we see the same qualitative behaviour: 
once the SMBH reaches its maximum, self-regulated mass, heating from the AGN overcomes the cooling losses and a strong energy-conserving outflows develops, 
with a very large mass outflow rate, close to 100~$M_\odot$/yr, but only for a short time. 
This short-lived outflow is enough to stop the accretion of fresh halo onto the disk (as seen on \autoref{fig:inflow_agn}).
The mass outflow rate at late time stabilises at the rather low value of 5~$M_\odot$/yr.
Comparing the two left panels on \autoref{fig:outflow_agn}, we see that the 50\,kpc mass outflow rate is higher than the 20\,kpc one, 
which is consistent with an outflow that sweeps the halo gas along its way.

\begin{figure*}
\begin{subfigure}{0.45\textwidth}
\centering
\includegraphics[width=\columnwidth]{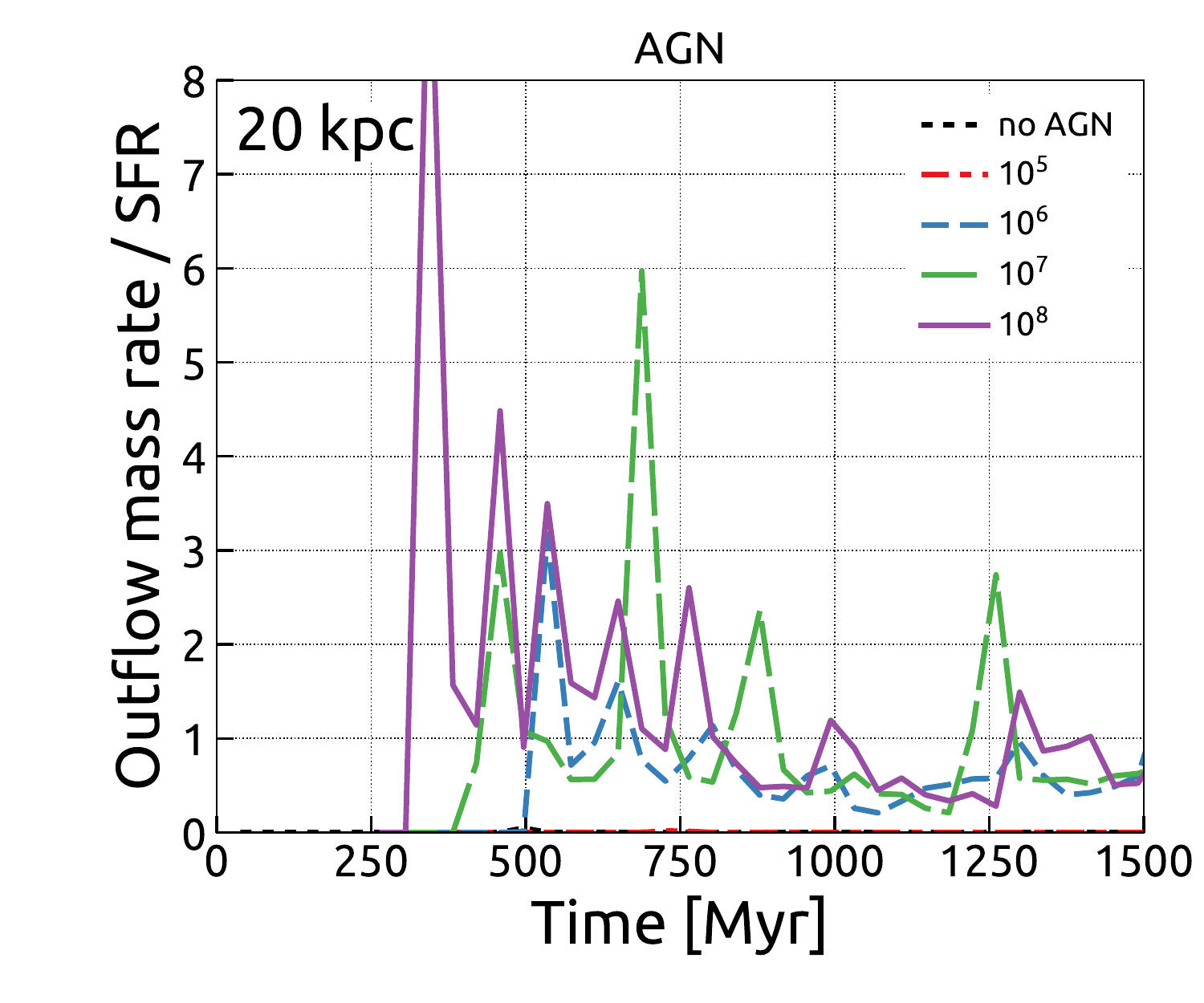}
\end{subfigure}
\qquad
\begin{subfigure}{0.45\textwidth}
\centering
\includegraphics[width=\columnwidth]{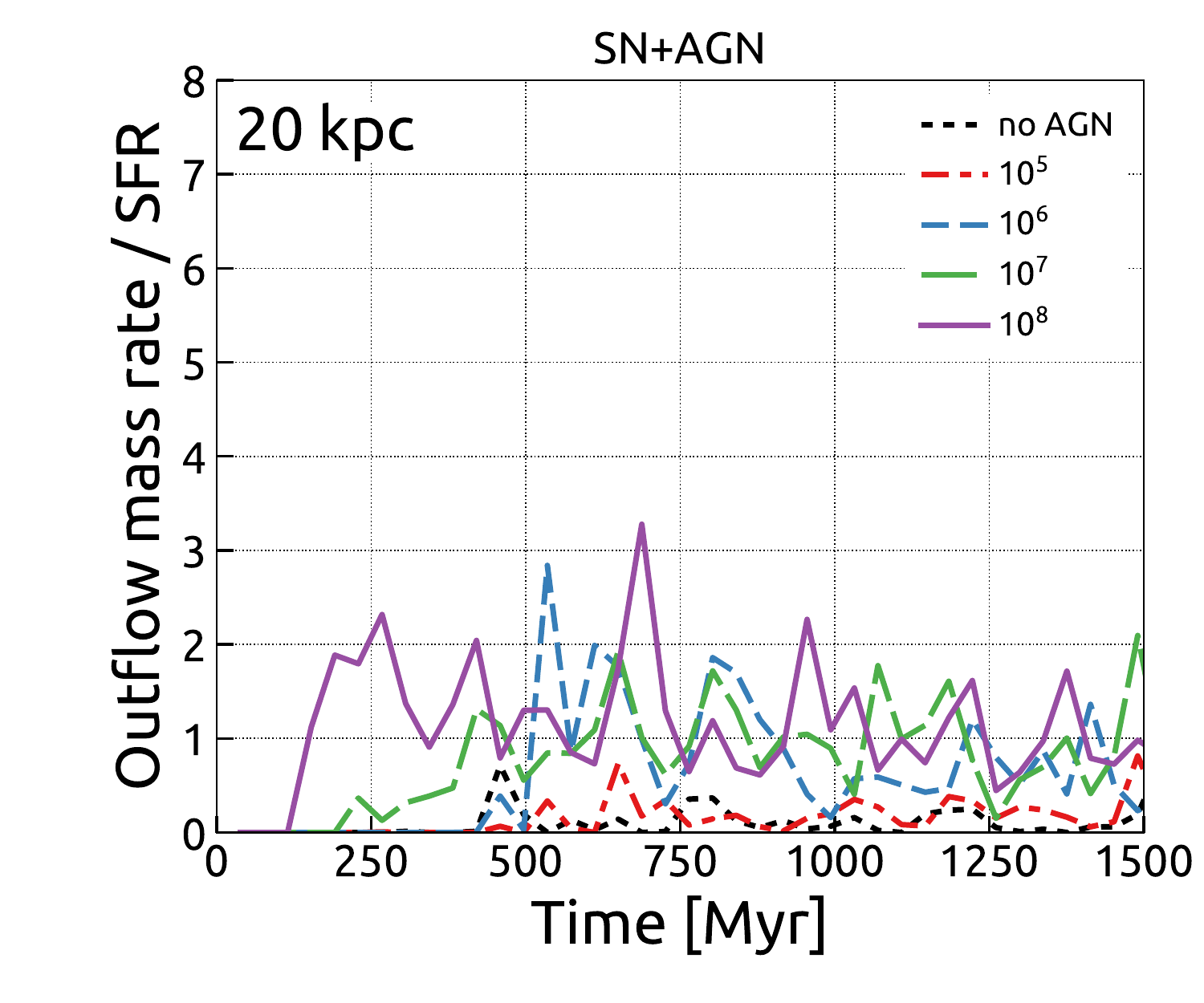}
\end{subfigure}
\\
\begin{subfigure}{0.45\textwidth}
\centering
\includegraphics[width=\columnwidth]{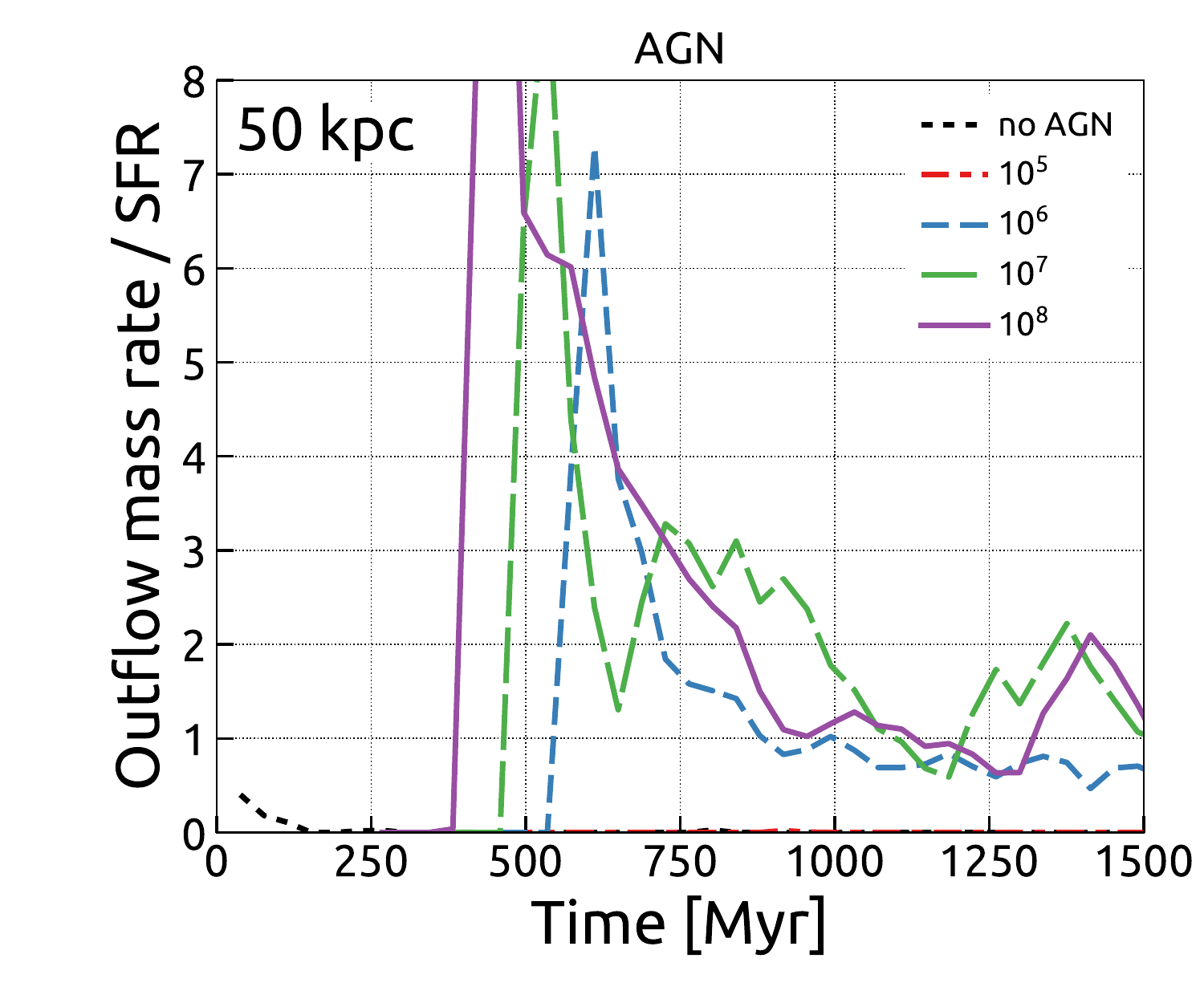}
\caption{AGN}\label{fig:loading_agn}
\end{subfigure}
\qquad
\begin{subfigure}{0.45\textwidth}
\centering
\includegraphics[width=\columnwidth]{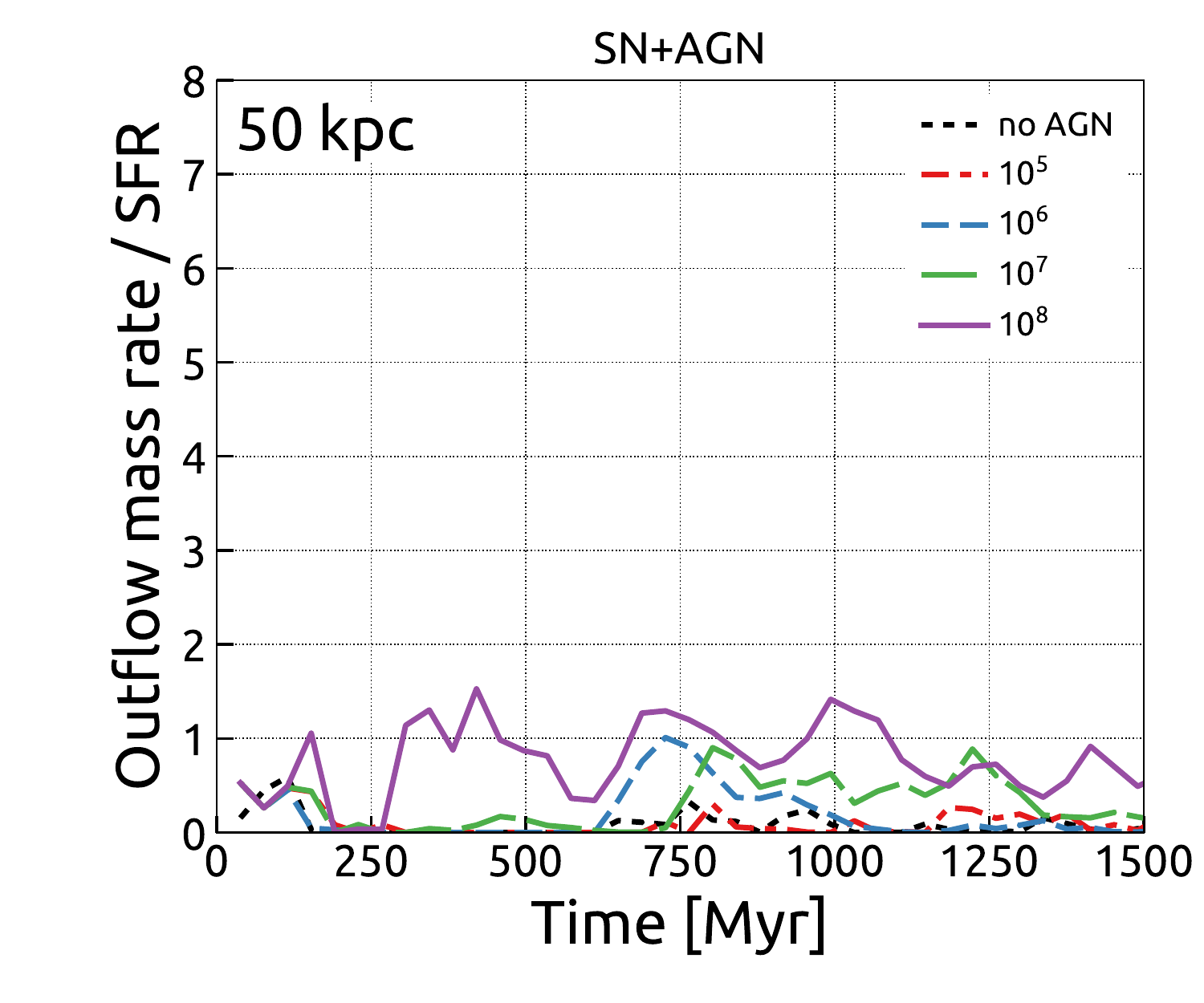}
\subcaption{SN+AGN}\label{fig:loading_sn_nsc}
\end{subfigure}
\caption{Mass loading parameter (outflow mass rate per star formation rate) in two sets of simulations - with AGN feedback only (left) and with both SN and AGN feedbacks (right) - for four different seed masses: $10^5$\,\msun{} - red (dash-dotted), $10^6$\,\msun{} - blue (short dashes), $10^7$\,\msun{} - green (long dashes), and $10^8$\,\msun{} - purple (solid); contrasted with runs without AGN feedback (black, dotted)). Left column: AGN only, right column: SN+AGN; top row: outflow measured through shell placed at 20 kpc from halo centre, bottom row: at 50 kpc.}
\label{fig:loading_20+50}
\end{figure*}

In \autoref{fig:loading_agn} we plot the mass loading factor for our five runs without SN feedback. 
This quantity is used to estimate if an outflow can efficiently regulate star formation.
The evolution of this mass loading factor can be divided in two periods: 1) an early epoch, when the SMBH just reached its maximum, self-regulated mass,
for which the mass loading factor is around $\sim 5$ and 2) a late epoch, when the mass loading factor falls down to $\sim 0.5$ ($\sim 1$) at 20\,kpc (50\,kpc).

The mass outflow rate measured in the simulation with only SN feedback is plotted as a black dotted curve on \autoref{fig:outflow_sn_nsc}. 
It rarely exceeds 5\,\msun/yr at 20 kpc from the centre, and is almost zero at 50 kpc. 
If AGN feedback is enabled, a strong and sustained outflow is produced, with a mass outflow rate around 20\,\msun/yr up to 50\,kpc.
It is worth noticing that in this case the mass outflow rate at late time is a factor of 5 larger than in the AGN-only simulations. 

\citet{ForsterSchreiber2014} find in their sample of massive galaxies at $z\sim 2$ ($\log(M_*/M_{\sun}) \geqslant 11$, comparable to our runs) 
clear signatures of AGN-driven outflows with $\dot{M}\sub{out}/\mathrm{SFR}\approx 3$, but ranging from $0.5$ to $15$, 
well within the range of the values produced by our simulations. 
We will compare our results to observations in greater detail in \autoref{sec:discussion}.
One effect that becomes apparent from a careful inspection of \autoref{fig:outflow_20+50} is the dependence of the outflowing mass to the initial seed mass. 
This weak effect is related to the synchronisation between the peak of the SF and the epoch when the SMBH reaches its maximum, self-regulated mass. 
It appears that the closer these events are to each other, the stronger is the outflow. 
This is due to the synchronisation of a strong SMBH accretion (due to the large reservoir of gas available) and a strong galactic fountain (in case SN feedback is present).

\begin{figure*}
\begin{subfigure}{0.45\textwidth}
\centering
\includegraphics[width=\columnwidth]{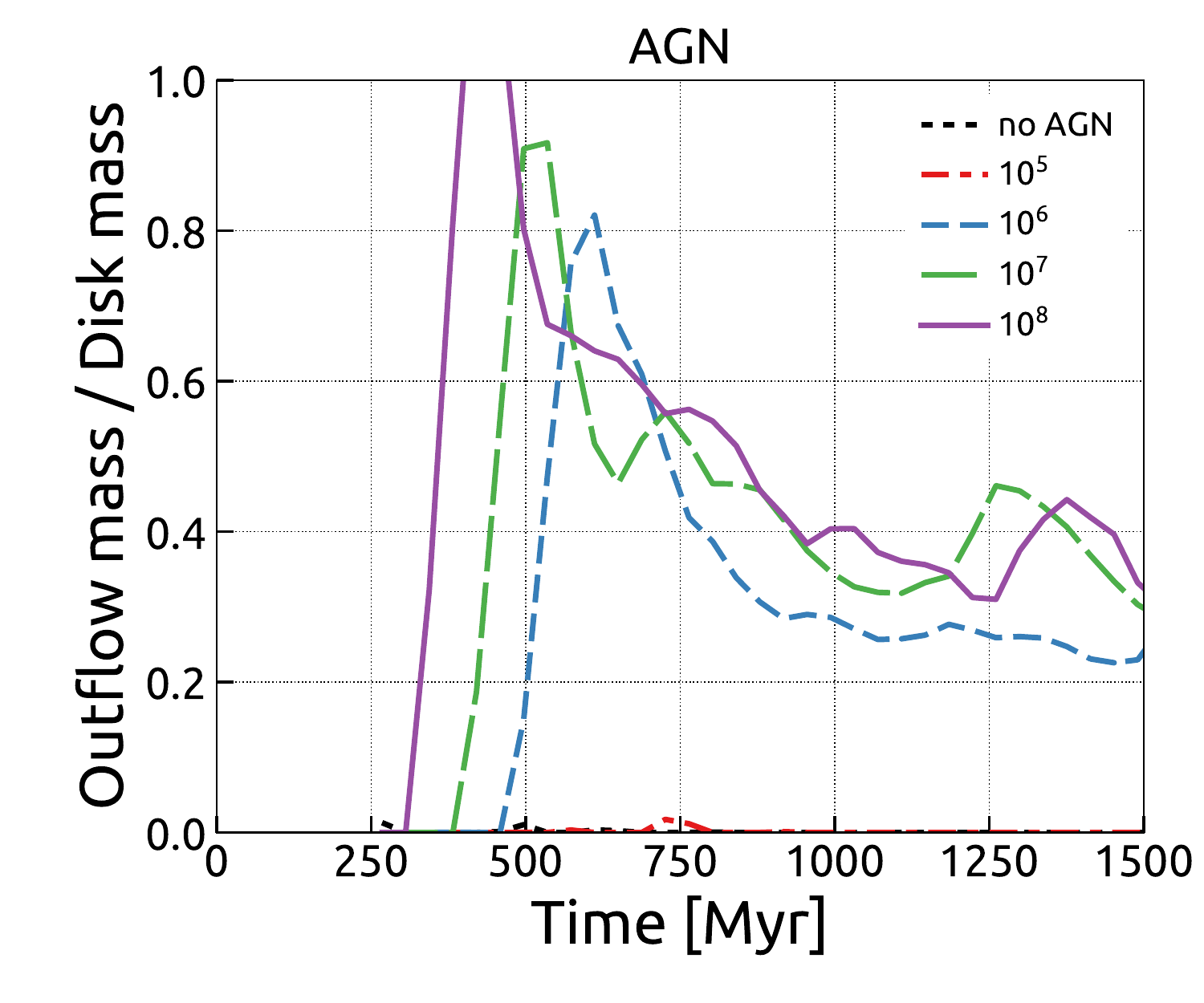}
\caption{AGN}\label{fig:mass_ratio_agn}
\end{subfigure}
\qquad
\begin{subfigure}{0.45\textwidth}
\centering
\includegraphics[width=\columnwidth]{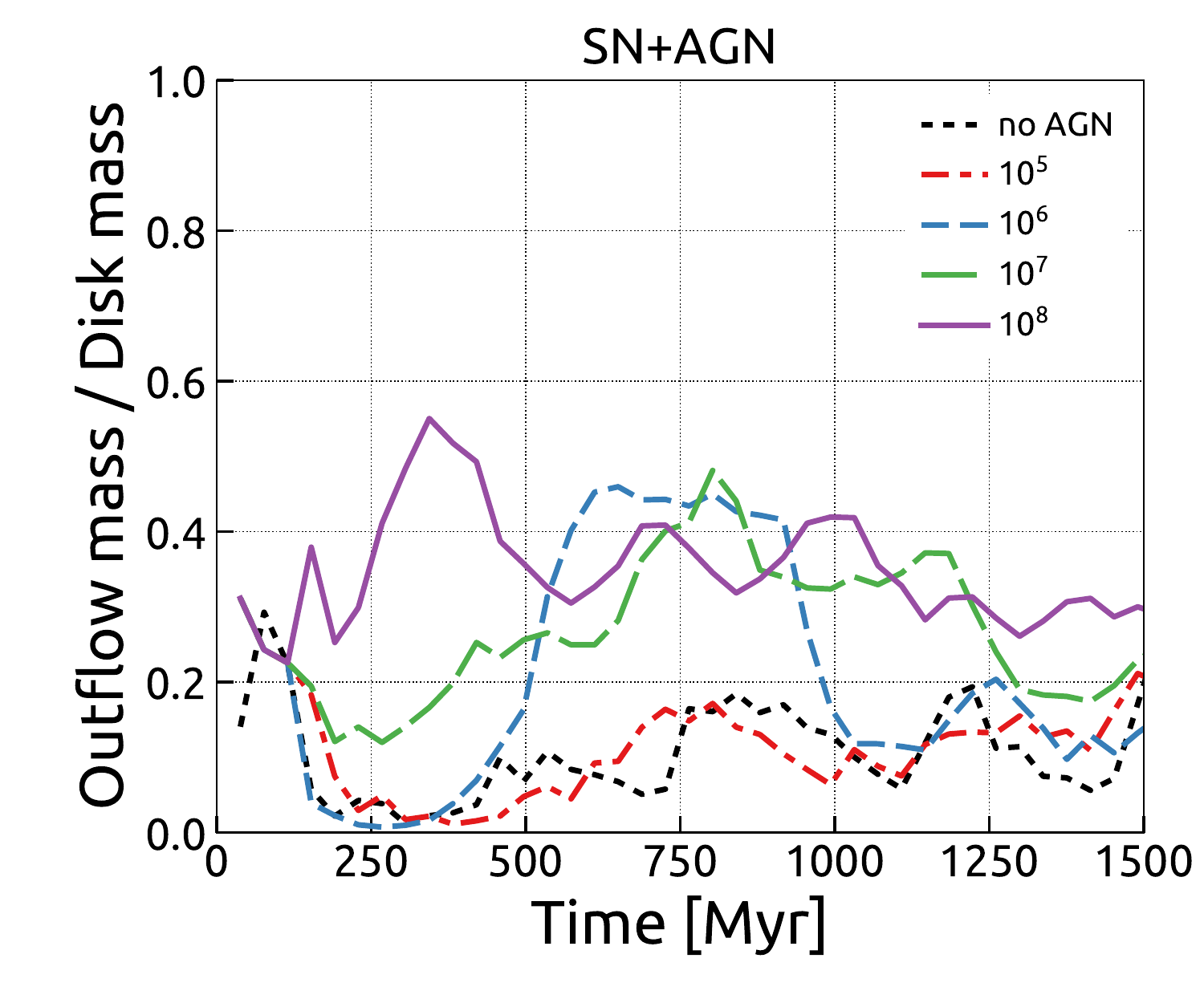}
\subcaption{SN+AGN}\label{fig:mass_ratio_sn_nsc}
\end{subfigure}
\caption{Time evolution of ratio between outflowing gas mass (green region in \autoref{fig:rhorho}, but excluding red) to gas mass in the disk ($\pm4$\,kpc from disk plane, red box in \autoref{fig:rhorho}) - for four different seed masses: $10^5$\,\msun{} - red (dash-dotted), $10^6$\,\msun{} - blue (short dashes), $10^7$\,\msun{} - green (long dashes), and $10^8$\,\msun{} - purple (solid); contrasted with runs without AGN feedback (black, dotted)). Left column: AGN only, right column: SN+AGN.}
\label{fig:mass_ratios}
\end{figure*}

Observationally, it is possible to characterise the outflows by comparing the amount of gas entrained in the outflow to the mass of the gas in a disk. 
In \autoref{fig:mass_ratios} we plot the ratio between the mass of all of the outflowing gas (selected as shown in \autoref{fig:rhorho}, 
i.e. out to 60\,kpc from the centre but excluding the galactic disk) and the mass of the gas contained in the disk ($\pm 4$\,kpc from the disk plane; the red region in \autoref{fig:rhorho}). 
This ratio reaches one when the outflow is the strongest (at early time) and falls down to 40\% at late time. 
In simulations with both feedback modes, AGN feedback is able to more than double the amount of gas entrained in the outflow, 
compared to the mass in the galactic fountain in the SN-only simulation. 
A larger SMBH seed leads to a more massive outflows , and earlier, an effect that we have already seen in \autoref{fig:loading_sn_nsc}.

\section{Outflow morphology}\label{sec:morphology}

In this Section we focus on the morphology of gas in our simulations at three different times that are fairly representative for different stages of the evolution. On \autoref{fig:rhorho} we have shown mass-weighted surface density projections for four different simulations (no feedback, AGN, SN and SN+AGN; top to bottom) at three different times (300, 750 and 1300 Myr; left to right). In the `no feedback' simulation there are no visible outflows and most of the gas is quickly consumed in star formation. In simulations with only AGN feedback there are no dense outflows, but halo gas is removed by a hot outflow, as discussed in \autoref{sec:outflows}. SN feedback on its own is able to produce gas that is violently evolving as a galactic fountain, propelled by continuous SN explosions in the galactic disk.
The morphology seen in the simulation with the cooperation between SN and AGN feedbacks is very different. Initially it resembles that of runs with SN only, as the SMBH requires time to grow to its self-regulation mass. Later on, AGN feedback launches dense and cold gas from the fountain to large radii.

\subsection{Temperature density diagram}\label{ssec:temp_rho}

In \autoref{ssec:phase_diag} we have discussed how the mass fraction changes as a function of density and radial velocity, while here we want to focus on the mass fraction as a function of density and temperature (see \autoref{fig:histograms}). The gas in the simulation with only AGN feedback is very diffuse and never cools below $10^6$\,K. This is in strong contrast with the temperatures found in the runs with SN feedback, where the outflow gas can cool to temperatures as low as few hundred Kelvin. This significant difference is explained if we recall that our simulations start with $Z\sub{ini}=0.05\,Z_{\sun}=0.001$ and that the only source of metal enrichment of the gas is via SN explosions. This leads to a lack of metals and associated cooling in the AGN-only simulation.

In the simulation with SN feedback a galactic fountain develops. The gas that is returning to the disk is cooler and denser (>1\,H/cc) than the outflowing gas, 
as revealed by the location of the densest gas on the phase space diagram.

If both feedback modes are included, we see very similar properties between the SN and the SN+AGN runs. 
In the latter, however, more {\it dense} gas is entrained in the outflow (cf. \autoref{fig:mass_ratio_sn_nsc}), 
that cools efficiently due to the higher metal enrichment. 
AGN feedback does not only accelerate the fountain gas, but also pushes it to larger radii, giving it more time to cool. 

\subsection{Radial profiles}\label{ssec:rad_profs}

\begin{figure*}
\centering
\includegraphics[width=\textwidth]{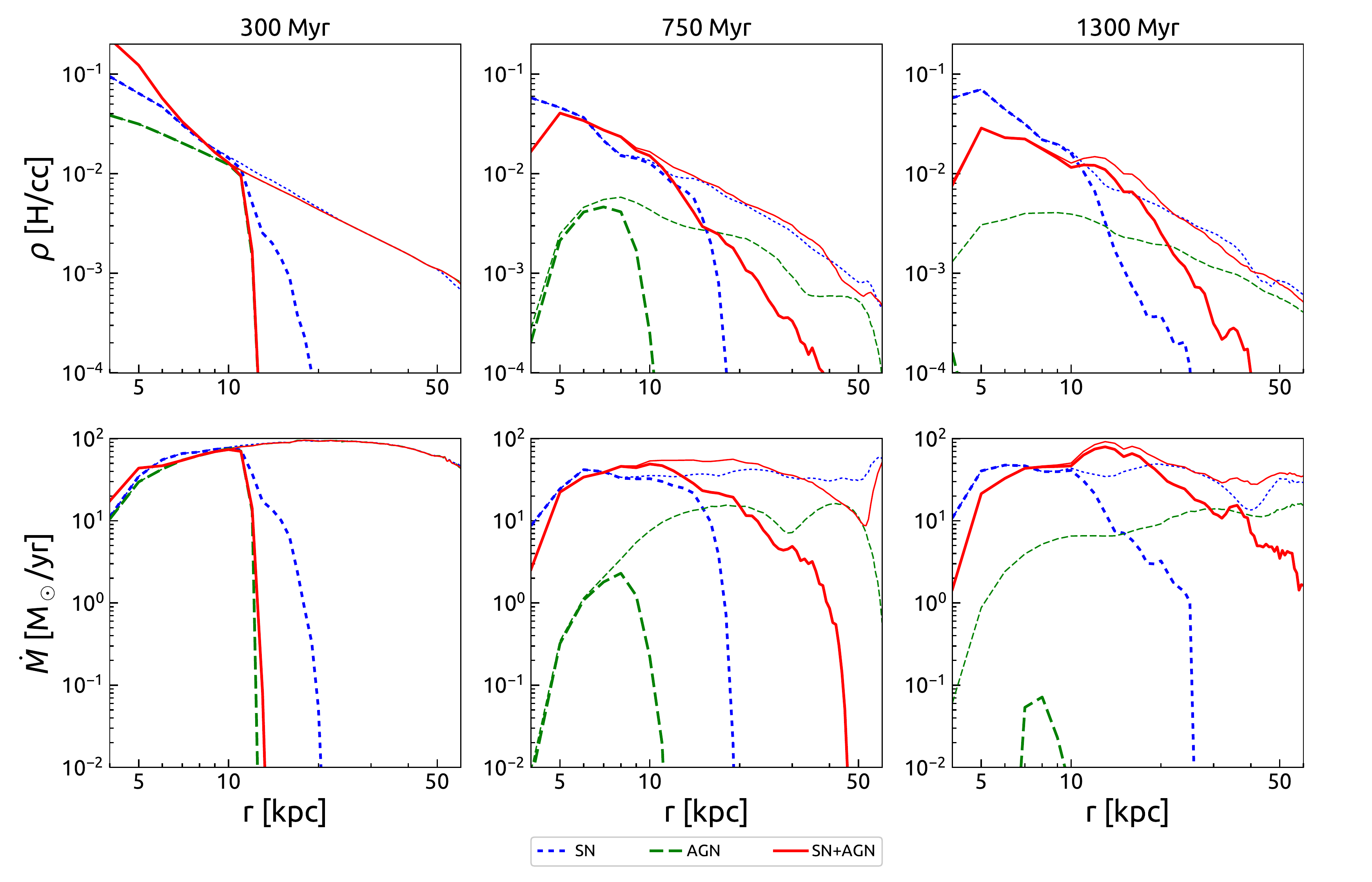}
\caption{Radial profiles of average density and average mass flow of \emph{inflowing gas} in a subset of simulations at three times: 300, 750 and 1300\,Myr (blue, dotted - SN feedback only; green, dashed - AGN feedback only; red, solid - SN+AGN). Thin lines mark all gas, while thick lines mark dense gas ($>0.01$\,H/cc). In all panels we consider only gas in the region outlined on \autoref{fig:rhorho}. \mseed{6} in simulations with AGN feedback modelled.}
\label{fig:gas_profiles_6_comp_in}
\end{figure*}

In \autoref{fig:gas_profiles_6_comp_in} we show the radial profiles of the average density (top row) and the average mass flow rate (bottom row) 
of the inflowing gas for three times representative for the halo evolution. 
All the gas is plotted with thin lines, while the dense gas,  defined as $n\sub{gas}>0.01$\,H/cc with thick lines. 
We use this threshold as it corresponds to self-shielded, neutral or possibly molecular gas (see below).
Initially (left column), the profiles are typical for an accretion flow from the extended halo, especially for the AGN run (plotted with green dashed line) 
which is not impacted by mixing from SN feedback. In runs with only AGN feedback we see that after the SMBH reaches its maximum self-regulated mass (middle and right panels), 
the gas has a significantly lower average density compared to runs where SN feedback is included. 
This means that the halo gas has been swept more efficiently in the AGN run than in the others. 

Interestingly, thanks to the effect of the combined feedback mechanisms (red lines), more dense gas is able to reach 50 kpc (middle column) and beyond (right column). 
Part of the outflow loses kinetic energy and starts falling back, thus also increasing the inflow rate. 
This is also reflected in the bottom row, where we plot average mass flow rate of the gas. 
As we discussed in \autoref{sec:sfr}, star formation is largely quenched, as AGN feedback \emph{prevents} gas from falling back onto the disk. 
This is in contrast with the SN+AGN run, in which an order of magnitude more gas is infalling onto the disk. 
We stress again the difference in metal enrichment between the runs - in simulations with SN feedback thanks to metal injection we observe more cooling and thus more dense gas. 
As a consequence, in the SN+AGN simulation, cooling boosts gas re-accretion and attenuate the effect of AGN feedback.
This explains why the SFR is not quenched as efficiently as in the AGN only case, 
as the mass inflow rate is an order of magnitude lower at all radii in the AGN only case compared to the other two runs.

\begin{figure*}
\centering
\includegraphics[width=\textwidth]{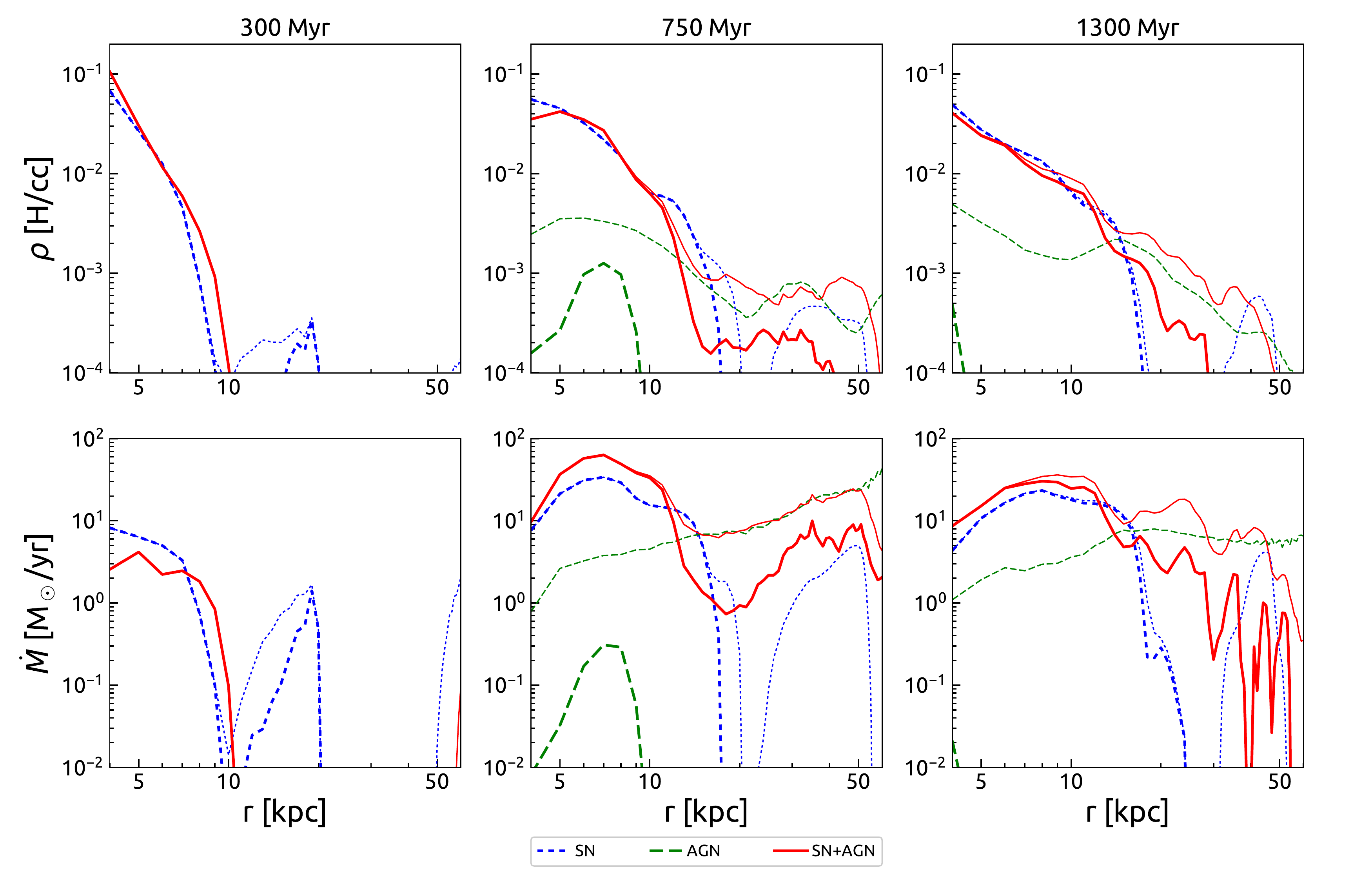}
\caption{Radial profiles of average density and average mass flow of \emph{outflowing gas} in a subset of simulations at three times: 300, 750 and 1300\,Myr (blue, dotted - SN feedback only; green, dashed - AGN feedback only; red, solid - SN+AGN). Thin lines mark all gas, while thick lines mark dense gas ($>0.01$\,H/cc). In all panels we consider only gas in the region outlined on \autoref{fig:rhorho}. \mseed{6} in simulations with AGN feedback modelled.}
\label{fig:gas_profiles_6_comp_out}
\end{figure*}

Turning our attention to the outflowing gas (\autoref{fig:gas_profiles_6_comp_out}, top row) we once again find lower density gas in the AGN-only simulation 
and higher density gas  in the SN-only run in the galactic corona (up to 20\,kpc from the centre). 
The absence of any outflowing gas at 300\,Myr in the AGN-only run is explained by the fact that the SMBH has not reached its maximum self-regulated mass yet. 
At later times, as seen in the bottom row of \autoref{fig:gas_profiles_6_comp_out}, very large quantities of dense gas are being expelled by combined AGN and SN feedbacks 
-- with $\dot{M}$ between 1 and 10\msun{}/yr, the latter value being larger than the inflow rate.
At 750\,Myr we clearly see that the mass outflow rate is rising with increasing radius, as it entrains more and more gas. 
In the case of the simulation with combined feedback mechanisms, the outflow is loaded with dense gas up to 50 kpc from the disk, as revealed by the thick line.

\begin{figure*}
\centering
\includegraphics[width=\textwidth]{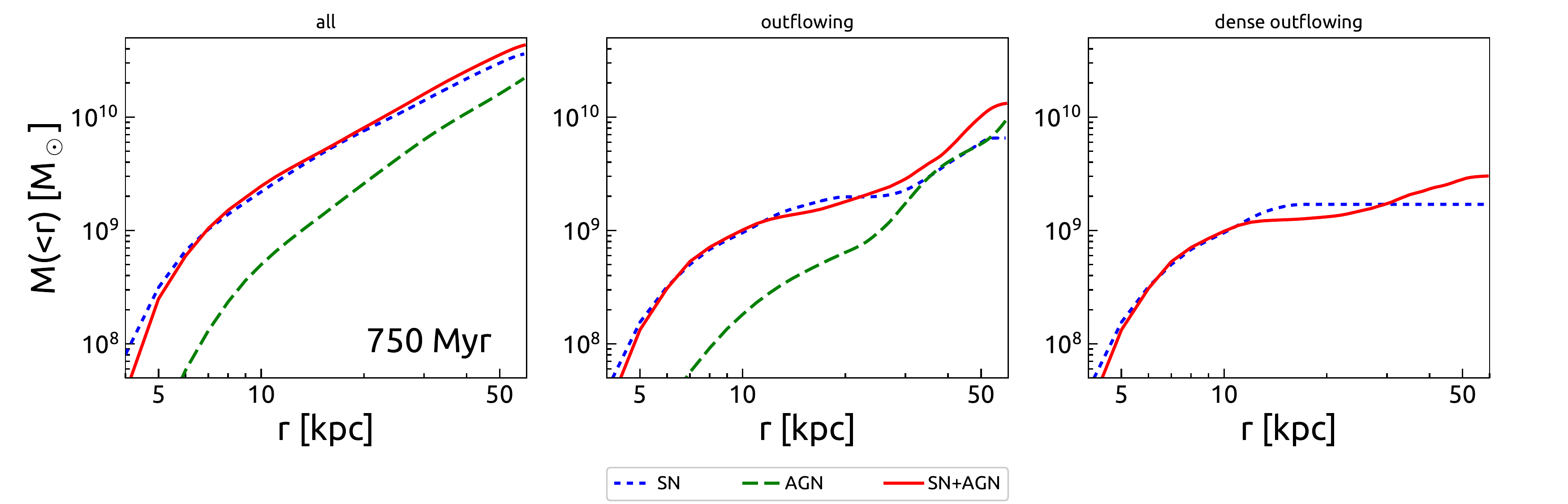}
\caption{Cumulative mass profile in a subset of our simulations at 750\,Myr. Left panel: all gas, middle panel: only outflowing gas, right panel: only dense, outflowing gas. Blue, dotted - SN feedback only; green, dashed - AGN feedback only; red, solid - SN+AGN. In all panels we consider only gas in the region outlined on \autoref{fig:rhorho}. \mseed{6} in simulations with AGN feedback modelled.}
\label{fig:gas_mass_profiles}
\end{figure*}

Another way to describe the gas distribution in our simulations is via cumulative mass profiles, which we show on \autoref{fig:gas_mass_profiles}. 
At 750\,Myr there is overall less gas at all radii in the simulation with AGN feedback only (green dashed line) compared to the other two runs. 
This means that more gas was removed and thus SF has been quenched more efficiently in the AGN-only case. 
If we now focus on the amount of the outflowing gas at $>30$\,kpc from the disk, we find that combined SN+AGN feedback is able to carry larger amounts of gas than each individual feedback mechanism on its own. Furthermore, limiting our analysis to only the dense outflowing gas (right column) we can make three important observations: 
1) in simulations with only AGN feedback there is no dense gas, 
2) the profile of SN feedback (blue dotted line) has no dense gas present beyond $\sim 15$\,kpc, while
3) we have twice more dense gas overall (up to 50\,kpc) in the SN+AGN simulation.

\begin{figure*}
\centering
\includegraphics[width=\textwidth]{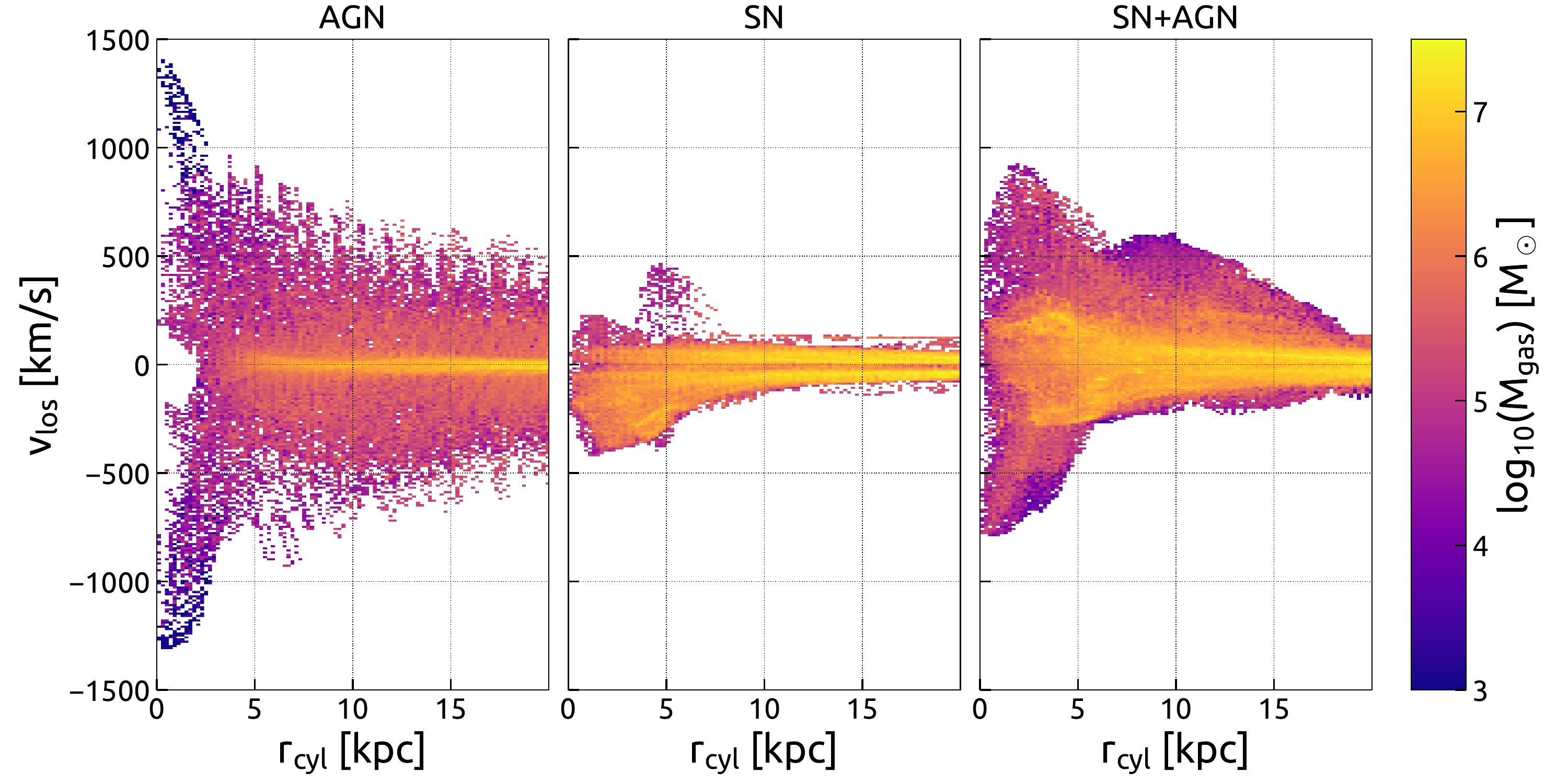}
\caption{Histograms of gas line-of-sight velocity when looking at the disk face on with respect to galactic radius $r\sub{cyl} $for three simulations with different feedback modes - AGN-only (left column), SN-only (centre), and SN+AGN (right column). We have selected only gas located at less than 60 kpc from the centre of the halo (green region on \autoref{fig:rhorho}) and excluded central disk of thickness 8 kpc (red rectangle on \autoref{fig:rhorho}). \mseed{6} in simulations with AGN feedback modelled.}
\label{fig:rxy_vr}
\end{figure*}

{On \autoref{fig:rxy_vr} we show mass-weighted histograms of the line-of-sight velocity $v\sub{los}$ at a given galactic radius $r\sub{cyl}$ (averaged in rings). 
In the AGN-only run the fastest moving gas is seen in the very centre; here gas is diffuse and hot. 
The fountain launched by the SN feedback shows lower velocities with a weaker radial dependency. 
The combination of the two feedback mechanisms is also centrally peaked in the same fashion as the AGN-only simulation, 
but this time it contains more dense and cold gas. 
It appears that our simulations seem to be in agreement with the radial dependency in observed galaxies in \citet{Genzel2014},
and with the simulations of star forming disks in \cite{Gabor2014}.}

\subsection{Evolution of baryonic mass}\label{ssec:evo_baryon}

\begin{figure}
\centering
\includegraphics[width=\columnwidth]{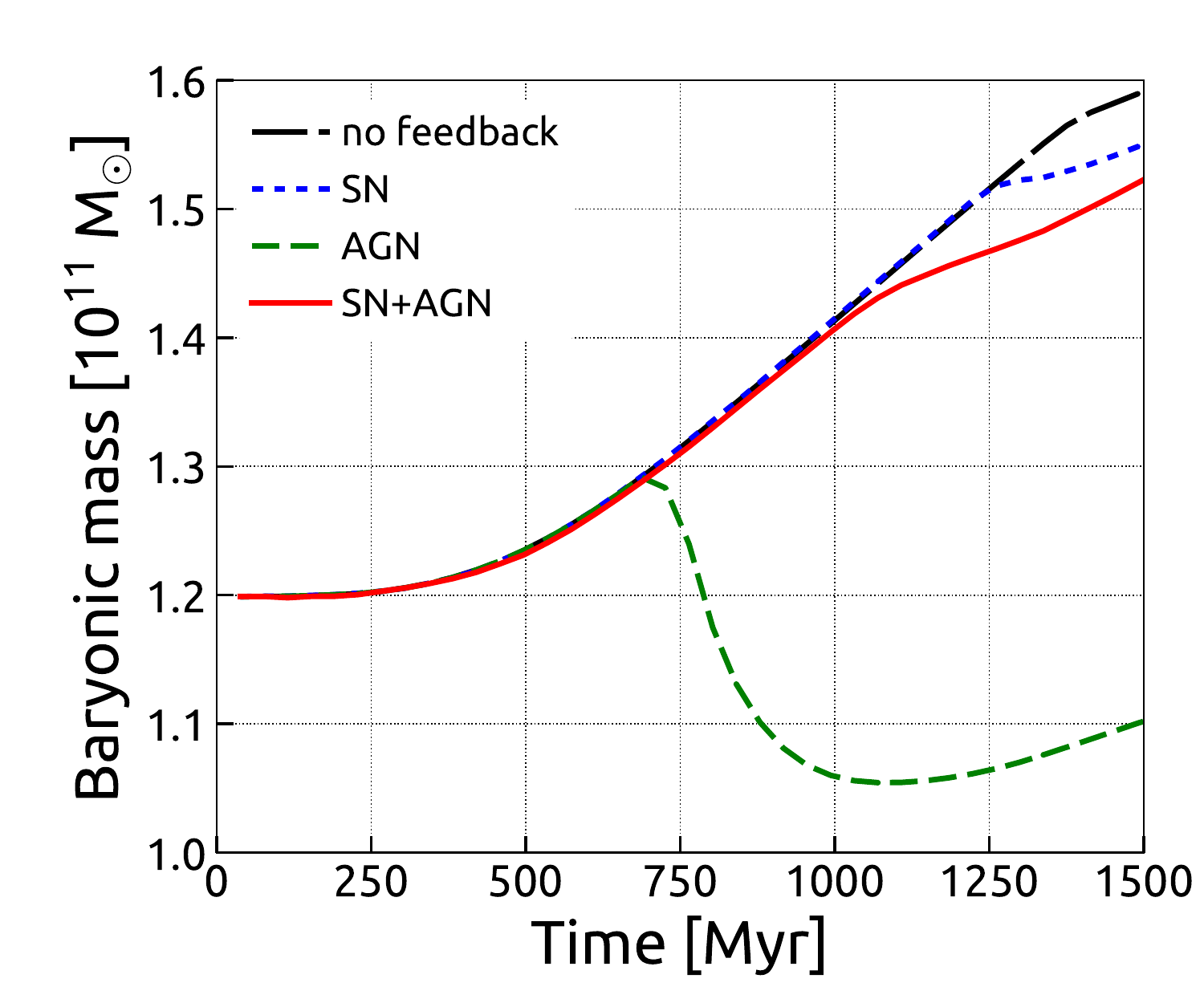}
\caption{Evolution of mass of baryons in  simulation without feedback (black, long dashes), SN-only (blue, dotted), AGN-only (green, short dashes) and SN+AGN (red, solid). We have selected all the gas, stars in inner 100\,kpc of the halo and the black hole (if simulated).}
\label{fig:baryon_frac}
\end{figure}

{The long-term secular processes can lead to slow depletion of the gas from the halo. In order to investigate if these processes take place in our simulations, we have measured the baryonic mass (stars, gas and a black hole) in four of our runs (no feedback, SN-only, AGN-only with \mseed{6}, and SN+AGN with \mseed{6}) within 100 kpc from the centre of the halo ($\sim0.5 R\sub{vir}$); see \autoref{fig:baryon_frac}. The baryonic mass in the no feedback run steadily increases with time and is always the highest among the four runs (reaching $1.6\times 10^{11}$\,\msun{} at 1500\,Myr). The baryonic content in the SN feedback run is reduced compared to the no feedback run ($1.55\times 10^{11}$\,\msun{}), suggesting that part of the halo gas can be removed by the long-term SN feedback. In the AGN-only run, in which we have initially the same evolution as in the SN-only case. Once the SMBH reaches its self-regulation mass, the AGN feedback is able to regulate the inflow via preventive feedback (to $1.06\times 10^{11}$ \msun{} at 1300\,Myr). By 1500 Myr some of the rate increases a bit, which suggests a traverse flow along the disk plane develops. In the SN+AGN run the baryonic mass increases with time, but at a rate few percent lower than that of SN-only run ($1.52\times 10^{11}$\,\msun{} at 1500\,Myr), suggesting that the AGN feedback is less efficient than in AGN-only run, but still reducing the baryonic mass within $\sim0.5 R\sub{vir}$.}


\section{Discussion}\label{sec:discussion}

\subsection{Molecular gas formation}

In order to compare our simulation results to observed molecular outflows at high-redshift, we would need to form molecular hydrogen self-consistently, 
which is far beyond the scope of this paper, and for which one would require much better spatial resolution.
As a consequence, we rely on a rather loose definition of ``dense gas'', adopting a density threshold $n_{\rm H}>0.01$\,H/cc. 
We would like to stress here that this value is often associated with the {\it self-shielding} density for neutral hydrogen \citep{Schaye2004, Aubert2010, Rahmati2013}.
Gas denser than this threshold will not be photo-heated and therefore can cool to arbitrary low temperature and ultimately form star. 
In order to strengthen our argument, we demonstrate now that this dense gas is indeed able to cool to low temperature, and is therefore a good candidate for 
molecular gas. 

In this paragraph, we will compare the cooling time of the dense gas in the outflow to the dynamical time in the halo.
The latter can be estimated as usual as
\begin{equation}
t\sub{dyn}=\sqrt{\frac{R\sub{200}^3}{GM\sub{200}}},
\label{eq:dynamical_time}
\end{equation}
\noindent where $R\sub{200}$ is a virial radius, $G$ is gravitational constant and $M\sub{200}$ virial mass of the halo. 
Using the particular values in \autoref{sec:setup}, we find that $t\sub{dyn}\approx 1.1$\,Gyr.
The cooling time can be estimated as
\begin{equation}\label{eq:cooling_time}
t\sub{cool}=\frac{\epsilon}{\dot{Q}} = \frac{3/2 nk\sub{B}T}{n^2\Lambda(T)},
\end{equation}
where $\epsilon$ is the internal energy of the gas, $\dot{Q}$ is the cooling rate, 
$k\sub{B}$ the Boltzmann constant, $n$ the gas number density and $\Lambda(T)$ the cooling function at temperature $T$. 

To compute the cooling time, we select all the gas from the regions marked on \autoref{fig:rhorho} that is outflowing and with a density greater than $0.01$\,H/cc.
We then compute the emission-weighted average density and temperature which turn out to be respectively $0.15$\,H/cc and $2\times10^6$\,K. 
Substituting these values in \autoref{eq:cooling_time} and using $\Lambda \simeq 2.3\times 10^{-23}$\,erg cm$^3$ s$^{-1}$, which is the value
of the cooling function at the average outflow temperature and metallicity (we find $Z \simeq 0.1$\,\zsun{}), we obtain $t\sub{cool}\approx 3.8$\,Myr, 
which is three orders of magnitude shorter than the halo dynamical time. 

One could argue that computing the dynamical time for the entire halo is not adequate, as we want to form molecular gas already in the galactic corona. 
If we define  the corona crossing time as $t\sub{cross}=R_{\rm corona}/v\sub{g}$, where we choose the size of the corona as $R_{\rm corona}\simeq 50\unit{kpc}$ 
and the maximum outflow velocity as $v\sub{g} \simeq1000\unit{km/s}$ (see \autoref{fig:histograms}), we get $t\sub{cross}\simeq50$\,Myr, still comfortably higher than the cooling time. 

This means that our dense outflowing gas will have enough time to turn molecular before traversing a significant fraction of the corona, not to mention the halo as a whole.  
To confirm our estimate, we analysed the temperature distribution within the outflow and found $\sim10^9$\,\msun{} of gas colder than $<10^4$\,K 
and $\sim10^8$\,\msun{} of gas colder than $<10^3$\,K. However, we would like to re-emphasise that we do not model explicitly molecular and radiation physics, 
thus this cold gas can only be interpreted as a tracer for the true molecular gas.

\subsection{Comparison to observations}

There is an increasing body of observational evidence of SMBH activity in galaxies at all redshifts. 
In the local Universe, AGNs are observed with fast, hot outflows ionising large quantities of gas in Seyfert 2 galaxies \citep[e.g.][]{Greene2011, Harrison2014, McElroy2015}. 
Powerful jets are also observed to produce large spherical cavities in galaxy clusters (see e.g. reviews by \cite{McNamara2007, Fabian2012, Heckman2014}). 

Observations at high redshifts ($z\approx1-3$) reveal massive molecular outflows in galaxies at the peak of star formation history, that also host bright quasars. 
{For example, the observations of $z = 2.3$ ultra-luminous infrared galaxy by \cite{George2014} find molecular outflow reaching velocities of 700\,\kms{}, somewhat lower than the estimated escape velocity of that object. Even more extreme outflow was recently observed by \cite{Feruglio2017} who have found a quasar with the outflow velocity of $\sim1340$\,\kms{}. In a very recent study, Chapman {et al.} (submitted) found a more extreme example of AGN-driven molecular outflows from the quasar HS1549+19 at $z = 2.84$.
They report observations of molecular gas with $v\sub{out}\sim1500$\,\kms{}. The stellar mass of the host galaxy ($\sim10^{11}$\,\msun{}) is significantly larger than in our case ($5\times10^{10}$\,\msun{}). 
HS1549+19 also hosts a more massive SMBH with $M\sub{SMBH}=4.6\times10^9$\,\msun{} compared to our case $2\times10^{8}$\,\msun{}. There are also multiple examples of cold outflowing gas seen at $z>5$ \citep[e.g.][]{Maiolino2012, Cicone2014}.
Not only cold dense outflows cane be seen in observations of massive galaxies. \cite{Genzel2014} presented a sample of $\sim30$ massive galaxies with broad nuclear emission and with a FWHM of $>450$\,\kms{} and reaching $\sim5000$\,\kms{}. These authors argue that stellar feedback can only account for outflows velocities up to $\sim200$\,\kms{}, which in turns mean that also the ionised outflows should be able to escape from the galaxy if they are propelled by AGN.} 

This observational picture agrees very well with our numerical results, as depicted in \autoref{fig:histograms}. 
The typical outflow velocities of the cold and dense gas component are only around $\sim500$\,\kms{} for the SN-only case, 
while for the SN+AGN model, they can reach much larger values around $1000$\,\kms{}, which is more than 5 times larger than $v\sub{circ}$.

In our previously paper \citepalias{Biernacki2017}, we have shown that the final mass of the black hole is related to the halo escape velocity and the size of the energy injection region by
\begin{equation}
v_{\rm esc} \simeq 750~{\rm km/s}~\left( \frac{M\sub{sink}}{10^8~M\sub{\odot}}\right)^{1/3} \left( \frac{R\sub{sink}}{100~{\rm pc}}\right)^{-1/3}\label{eq:csad}
\end{equation}
This suggests that, if the energy driving the outflow is deposited in a region similar to our simulations, the outflow velocity can be up to 2.5 times larger than in our simulation, 
easily reaching the observed value. This means that the escape velocity of HS1549+19 should also be close to 1500\,\kms{}, which is a rather extreme value.
Note that \cite{Nesvadba2011} also reported earlier the discovery of two $z\geq3.5$ quasars with large-scale outflows and FWHM velocities up to $5000$\,\kms{}. 

In \citet{ForsterSchreiber2014}, a sample of massive $z\sim 2$ galaxies observed with SINFONI has been presented, with stellar masses and mass loading factors similar to our simulations. 
They have speculated that nuclear outflows driven by AGN feedback is probably a general characteristic of massive galaxies at the peak of their star formation history (i.e. $z\sim 2$). 
Our simulation clearly confirm this picture and reveal the physical mechanism that powers these massive outflows, namely the combination of efficient SN feedback in conjunction with a powerful AGN. 
At even higher redshifts than discussed here, evidence of quasar-driven outflows does also exist. 
Very massive, gas-rich galaxies at $z\approx5-7$ seem to be excellent hosts for both efficient SMBH-fuelling and efficient SF through dense gas clumps, explaining recent observations \citep{Aalto2012, Cicone2014, Cicone2015} with outflows reaching $1400$\,\kms{}. 

\subsection{Comparison to previous simulations}

Feedback from SMBH has been invoked a  long time ago to explain the luminosity function of galaxies at the high mass end \citep[e.g.][]{Silk1998}.
It has been included since then in both semi-analytical and numerical models of galaxy formation, and acts as the main mechanism leading to the so-called quenching of SF in massive galaxies 
with $M\sub{halo}>10^{12}$\,\msun{} \citep{diMatteo2005, Croton2006, Hopkins2006, Somerville2008,Dubois2010,Teyssier2011, Fabian2012,Feldmann2016}. It is however still unclear how this process occurs in details. 

Results from recent large cosmological simulations \citep{Vogelsberger2014, Schaye2015, Dubois2016} show that AGN feedback is a necessary ingredient of a successful galaxy evolution model. 
On the other hand, in very high resolution simulations of isolated galactic discs, \cite{Gabor2014} and \cite{Roos2015} have demonstrated that AGN feedback has very little effect on the SF within the disk. 
The SMBH could in principle release as much as $10^{59}$\,erg of energy ($E\sub{SMBH}=0.1M\sub{SMBH}c^2$), 
largely exceeding binding energy of the galaxy ($E\sub{gal}\approx M\sub{gas}\sigma^2$, where $\sigma$ is the velocity dispersion).

As explained by \cite{Gabor2014} and \cite{Roos2015}, this naive expectation turned out to be wrong
for mainly two reasons: (1) the energy is deposited in a very small region around the SMBH and 
(2) this energy quickly escapes the nuclear region,
either buoyantly (if the gas is hot) or ballistically (if the gas is cold), without affecting the disk significantly.
A clear result of our simulation suite is that indeed AGN feedback does not affect the SF in the disk via the \emph{ejective} mode, 
but it does \emph{prevent} gas from inflowing from the corona or from the larger scale halo.  
This result has been confirmed phenomenologically by recent cosmological simulations \citep{Vogelsberger2014, Schaye2015, Dubois2016}. 
It is therefore crucial to include gas infall from either a cooling halo or a realistic cosmological environment to truly assess the effect of AGN feedback 
on the SF history of the simulated galaxy.

Another interesting aspect of our simulation is the emergence of a high velocity molecular (or at least dense and cooling) outflow. 
A recent cosmological simulation by \cite{Costa2015} also obtained such an outflow with dense and cold gas, with velocities reaching $1400$\,\kms{}. 
They argue that these dense outflows emerged through the interaction of dense cold filaments around the galaxy and the hot AGN-driven outflow from the SMBH.
Similar results were obtained by \cite{Prieto2017}, who showed that the SN-driven galactic fountain could also play a role in their high-$z$, clumpy galaxies, 
fed by cold gas-rich filaments. {These authors concluded that the origin of the cold gas in their outflows is due to the rarefaction of gas by SN and further push by AGN.}
In the case of our idealised cooling halo simulations, we can ascertain that the gas propelled by AGN feedback originates from the galactic fountain and is therefore metal-enriched, 
rather than the pristine gas from cold streams as in simulations of \cite{Costa2015} and possibly in \cite{Prieto2017}. 
We also note that in a very recent cosmological simulations, \citet{Pontzen2017} have also observed AGN feedback launching a low-density and high-velocity outflow sweeping the SN-driven fountain gas.

{The SN feedback model used in our simulations is far from being realistic. The physical processes involved in launching of the outflows are modelled phenomenologically with a subgrid model of delayed cooling or not modelled at all (e.g. stellar winds). Naturally, different implementations of SN feedback can lead to drastically different images of a galaxy. We can imagine two opposite results: 1) SN feedback that is too weak to push gas, that would result in a thin gas disk and 2) very strong SN feedback, that would blow all the gas away \citep[e.g.][]{Bournaud2014, Hopkins2014}. In our setup we have aimed to achieve a qualitative result that produces galactic fountain seen in some of the observations. We caution that our predictions are qualitative, but nevertheless allow us to explore the effects of coupling between SN feedback and AGN feedback.}


\section{Summary}\label{sec:summary}

In this work, we have analysed the effect of our new SMBH feedback recipe presented in \citetalias{Biernacki2017} on the quenching of star formation and on the launching of gas outflows. 
Thanks to high-resolution simulations of an isolated, gas-rich cooling halo, we are able to reproduce a realistic galactic environment with gas inflow, 
while resolving the ISM structure in the disc with a resolution of 100~pc. 

The feedback mechanisms included in our simulations have led to the launching of strong outflows with different characteristics. 
Purely AGN-driven outflows are hot and diffuse, and only sweep up gas in the outer galactic halo. 
These outflows are launched when the SMBH reaches its maximum, self-regulated mass. 
These hot outflows are affecting the SF history of the galaxy by preventing fresh gas from being accreted in the disk or ejected gas from falling back to the disk. 
Without SN metal enrichment, pure AGN-driven outflows cannot cool and form dense gas that could become molecular. 

In simulations with only SN feedback, we observe the formation of a dense galactic fountain, that can be characterised by cold gas with moderate velocities, bound to the galactic disk. 
In simulations with both feedback models together, a clear synergy was revealed -- SN feedback creates a galactic fountain with dense gas clumps and AGN feedback launches a low-density, hot outflow that sweeps the galactic corona, pushing the dense clumps to large distances. 
The resulting outflows is much more sustained and carries away a larger amount of mass. 
Metal enrichment from SN feedback promotes more cooling in the corona and, as a consequence, more gas can fall back onto the disk.

The mass loading factor of the simulated outflows are found to be close to unity, as seen in many observations. 
The analysis of the kinematic properties of the outflows reveals that the AGN is the main source of energy for the dense, molecular outflows. 
We have shown that these massive outflows can quench star formation in galaxies. This does not proceed via \emph{ejective} feedback, but via \emph{preventive} feedback, 
cutting the supply of fresh gas into the disk. 

One requirement we find is a delicate synchronisation between an active star formation phase, that can trigger the formation of galactic fountain, 
and a central SMBH reaching its maximum mass, that can trigger the formation of a fast, AGN-driven, hot outflow.
We speculate here that the simultaneity of these two conditions -- sustained star formation and the SMBH reaching its final mass -- can happen immediately after a ``wet compaction'' event, as described in e.g. \cite{Dekel2014}. It is sometimes related to the bulge formation epoch \citep[e.g.~][]{Dubois2013a}. 

In summary, we argue that SF can be quenched by AGN through \emph{preventive} feedback. 
We have also shown that fast and dense outflows can arise when SN and AGN feedbacks act in tandem, and that this happens at a very specific epoch of the galaxy life,
when star formation is still active while the SMBH reaches its maximum mass. 

\section*{Acknowledgements}
	We would like to thank the referee, Frederic Bournaud, for valuable comments that improved the quality of this work and presentation of the results. We appreciate helpful conversations with Arif Babul, Pedro R. Capelo, Tiago Costa, Alireza Rahmati, Valentin Perret, Oliver Hahn and Chiara Feruglio. Simulations performed for this work were executed on zBox4 at University of Zurich and on Piz Dora/Piz Daint at Swiss Supercomputing Centre CSCS in Lugano. This work made use of \textsc{pynbody} package \citep{pynbody}. PB benefitted greatly from participation in the BIRS-CMO workshop ``Computing the Universe: At the Intersection of Computer Science and Cosmology''.

\bibliographystyle{mnras}
\def\apj{ApJ}
\def\apjs{ApJS}
\def\apjl{ApJL}
\def\aj{AJ}
\def\mnras{MNRAS}
\def\aap{A\&A}
\def\nat{Nature}
\def\pasj{PASJ}
\def\prd{PRD}
\def\physrep{Physics Reports}
\def\jcap{JCAP}
\bibliography{feedback}

\end{document}